\documentclass[12pt,preprint]{aastex}
\usepackage{amsmath}

\begin{document}
\title{Observations and Magnetic Field Modeling of a Solar Polar Crown Prominence}

\author{Yingna Su\altaffilmark{1}, Adriaan van Ballegooijen\altaffilmark{1}}
\altaffiltext{1}{Harvard-Smithsonian Center for Astrophysics, Cambridge, MA 02138, USA.}  
\email{ynsu@head.cfa.harvard.edu}

\begin{abstract}

We present observations and magnetic field modeling of the large polar crown prominence that erupted on 2010 December 6. Combination of SDO/AIA and STEREO$\_$Behind/EUVI allows us to see the fine structures of this prominence both at the limb and on the disk. We focus on the structures and dynamics of this prominence before the eruption. This prominence contains two parts: an active region part containing mainly horizontal threads, and a quiet Sun part containing mainly vertical threads. On the northern side of the prominence channel, both AIA and EUVI observe bright features which appear to be the lower legs of loops that go above then join in the filament. Filament materials are observed to frequently eject horizontally from the active region part to the quiet Sun part. This ejection results in the formation of a dense-column structure (concentration of dark vertical threads) near the border between the active region and the quiet Sun. Using the flux rope insertion method, we create non-linear force-free field models based on SDO/HMI line-of-sight magnetograms. A key feature of these models is that the flux rope has connections with the surroundings photosphere, so its axial flux varies along the filament path. The height and location of the dips of field lines in our models roughly replicate those of the observed prominence.  Comparison between model and observations suggests that the bright features on the northern side of the channel are the lower legs of the field lines that turn into the flux rope. We suggest that plasma may be injected into the prominence along these field lines. Although the models fit the observations quiet well, there are also some interesting differences. For example, the models do not reproduce the observed vertical threads and cannot explain the formation of the dense-column structure.

\end{abstract}

\clearpage


\section{INTRODUCTION}

Solar prominences are relatively cool structures embedded in the
million-degree corona \citep{1985SoPh..100..415H, 2010SSRv..151..243L, 2010SSRv..151..333M}. 
In H$\alpha$ when viewed above the solar limb, prominences appear as bright structures against the
dark background, but when viewed as ``filaments'' on the solar disk
they are darker than their surroundings. We will use the terms ``filament" and ``prominence" 
interchangeably in general. A filament is formed in a filament channel, which is defined as a region in the chromosphere
surrounding a polarity inversion line (PIL) where the chromospheric H$\alpha$ fibrils are aligned with the PIL \citep{1971SoPh...19...59F,1998ASPC..150..257G}.
Filaments can be found inside activity nest consisting of multiple bipolar pairs of spots (``active region filaments"), at the border of active regions
(``intermediate filaments"), and on the quiet Sun (``quiescent filaments"), including the polar crown.

Filaments typically consist of  three structural components: a spine, barbs, and two extreme ends \citep{1998SoPh..182..107M, 2008ASPC..383..235L, 2011SSRv..158..237L}. The spine runs horizontally along the top of the filament, although there may be sections along the filament
where the spine is nearly invisible. The barbs protrude from the side of the filament and when observed closer
to the limb, the barbs are seen to extend down from the spine to the chromosphere below. The ends, also called ``legs", may be a collection
of threads that appear to terminate at a single point or at multiple points. When a quiescent filament is viewed with high resolution on the solar disk, H$\alpha$ observations indicate that each of these three structural components consist of thin thread-like structures \citep{2008ASPC..383..235L, 2011SSRv..158..237L}. In active regions the thin filament threads often appear to be mostly horizontal \citep{2007Sci...318.1577O, 2008ASPC..383..235L}, while quasi-vertical threads (``hedgerows") are often seen in the quiescent prominences \citep{2008ApJ...676L..89B, 2010ApJ...716.1288B, 2008ApJ...689L..73C, 2008ApJ...686.1383H}.  Prominence plasma is highly dynamic, exhibiting horizontal \citep{2008ApJ...689L..73C} or vertical \citep{2008ApJ...676L..89B}  flows. The flows reported so far are either unidirectional \citep {2008ApJ...676L..89B, 2008ApJ...689L..73C} or bi-directional  \citep{1998Natur.396..440Z, 2003SoPh..216..109L}. The latter is known as counter-streaming.

The magnetic field plays a primary role in filament formation, stability, and eruption \citep{Priest1989, Tandberg-Hanssen1995, 2010SSRv..151..333M}. However, the magnetic structure of prominences is still not fully understood, with many observations and theoretical models differing on the exact nature of 
the magnetic field. Various models for prominence magnetic structure can be summarized as follows.
In most of the models, filament plasmas are assumed to be located near the dips of magnetic field lines.
These models can be classified as ``sheared arcade model'' \citep{1994ApJ...420L..41A, 2000ApJ...539..954D, 2002ApJ...567L..97A} and ``flux rope model", \citep{1974A&A....31..189K, 1983SoPh...88..219P, 1989ApJ...344.1010P, 1989ApJ...343..971V, 1994SoPh..155...69R,
1995ApJ...443..818L, 1998A&A...335..309A, 2001ApJ...560..476C, 2004ApJ...612..519V, 2006ApJ...641..590G,
2008SoPh..248...29D}. In these models, the prominence plasmas are embedded in a  large sheared arcade or a flux rope
that lie horizontally above the PIL. Flux rope models may be split into two categories: weakly twisted
flux rope and highly twisted flux rope. Weakly twisted flux rope models
are similar to sheared arcades with one key difference as argued by \citet{2010SSRv..151..333M}. For flux rope models
the flux rope and overlying arcade are independent flux systems with a separatrix surface between them. In contrast,
for a sheared arcade model only a single flux system exists. In addition to the above mentioned 3D global models, 
there are also two other dip models which focus on the local support of vertical threads in hedgerow prominence, i.e., ``sagged horizontal fields
model" \citep{2008ApJ...689L..73C, 2010ApJ...714..618C} and ``tangled magnetic field  
model" \citep{2010ApJ...711..164V}. In the earlier model, the vertical threads are stacks of plasma supported against gravity by the sagging of initially horizontal magnetic field lines. In the latter model, tangled fields exist in a vertical current sheet of small width that is confined
by the vertical fields on either side of the sheet. Neither of these models describes the global 3D topology of the magnetic field supporting the filament.
In contrast to models mentioned above, the ``field aligned thread model" developed by \citet{2008ASPC..383..235L, 2011SSRv..158..237L} does
not contain dips. This model is a 3D empirical magnetic model based on high-resolution H$\alpha$ observations and is based on the assumption that the observed fine scale structures are parallel to the magnetic field. In this model, the filament plasma
is located on magnetic arches that are highly sheared in the direction along the PIL. Some field lines run along the entire length
of the filament and outline the filament ``spine". Other shorter ones run partially along the spine, but spread out from it
and connect down to minority polarity elements on either side of the PIL. These shorter structures represent the filament barbs.
In this case, the plasma is supported by MHD waves. However, relatively high frequencies and wave amplitudes are required,
and it is unclear why such waves would not lead to strong heating of the prominence plasma.

The purpose of the present paper is to analyze observations of a quiescent prominence, and to develop 3D models of its magnetic structure. Filaments in active regions have been modeled extensively in the past. Several authors have developed non-linear force-free field (NLFFF) models which are based on magnetic observations. For example, \citet{2007A&A...468..701R} construct NLFFF models by ``extrapolating" observed photospheric vector fields into the corona \citep[also see][]{2002A&A...392.1119R, 2010ApJ...715.1566C}. van Ballegooijen (2004) developed an alternative method for constructing NLFFF models of filament flux ropes. The method involves inserting a flux rope into a potential field based on an observed photospheric magnetogram, and then evolving the field in time to an equilibrium state using magneto-frictional relaxation. This method has been used by \citet{2008ApJ...672.1209B} and \citet{2009ApJ...704..341S, 2009ApJ...691..105S, 2011ApJ...734...53S, 2012ApJ...746...81A} to study active regions with filaments, and by \citet{2009ApJ...703.1766S, 2012ApJ...744...78S, 2012ApJ...750...15S} to study the evolution of soft X-ray sigmoid. The papers mentioned above suggest that the observed active region filaments can be well explained by a flux rope model. In this paper we apply this flux rope insertion method to a quiescent prominence for the first time.

\section{Observations}

\subsection{Data Sets and Instruments}

A large polar crown prominence was observed in the period December 1--6  in 2010, by the Atmospheric Imaging Assembly \citep[AIA, ][]{2012SoPh..275...17L} aboard the $\emph{Solar Dynamics Observatory}$ (SDO), as well as the STEREO$\_$B (Behind)/EUVI \citep{2004SPIE.5171..111W, 2008SSRv..136...67H}. Synoptic observations by the X-Ray Telescope \citep[XRT, ][]{2007SoPh..243...63G, 2008SoPh..249..263K} aboard $\emph{Hinode}$ \citep{2007SoPh..243....3K} and H$\alpha$ observations by the Kanzelh$\ddot{o}$he Solar Observatory (KSO) are also included in the study. The photospheric magnetic field information is provided by the Helioseismic and Magnetic Imager \citep[HMI, ][]{2012SoPh..275..229S} aboard SDO and SOLIS (Synoptic Optical Long-term Investigations of the Sun).  

The large prominence erupted around 14:18 UT on 2010 December 6. This prominence eruption is associated with a CME, which has a linear speed of 538 km s$^{-1}$, according to the SOHO LASCO CME catalog\footnote{http://cdaw.gsfc.nasa.gov/CME$\_$list/}. Detailed studies on the prominence eruption are presented by \citet{Su2012, Thom2012}.  In the current paper, we present the structure and dynamics of the prominence before and after the eruption.

\subsection{Structure and Dynamics of the Prominence and Surroundings}

\subsubsection{Prominence Structure and Dynamics before the eruption}

\begin{figure} 
\begin{center}
\epsscale{1.0} \plotone{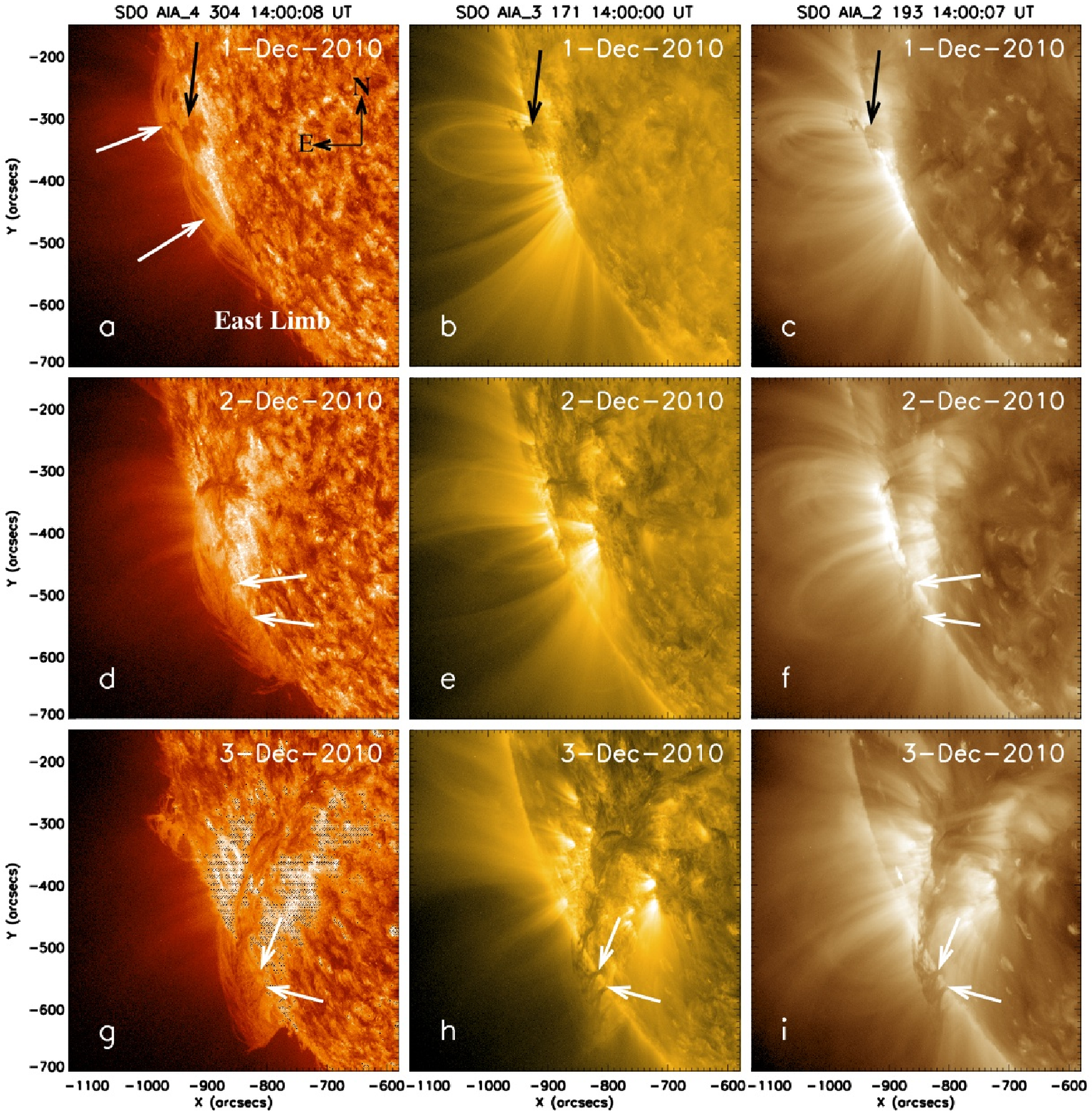}     
\end{center}
\caption{SDO/AIA observations of the quiescent prominence at 14:00 UT on 2010 December 1 (top row),  December 2 (middle row), and December 3 (bottom row). (A color version of this figure is available in the online journal.) \label{fig1}}
\end{figure}

\begin{figure} 
\begin{center}
\epsscale{1.0} \plotone{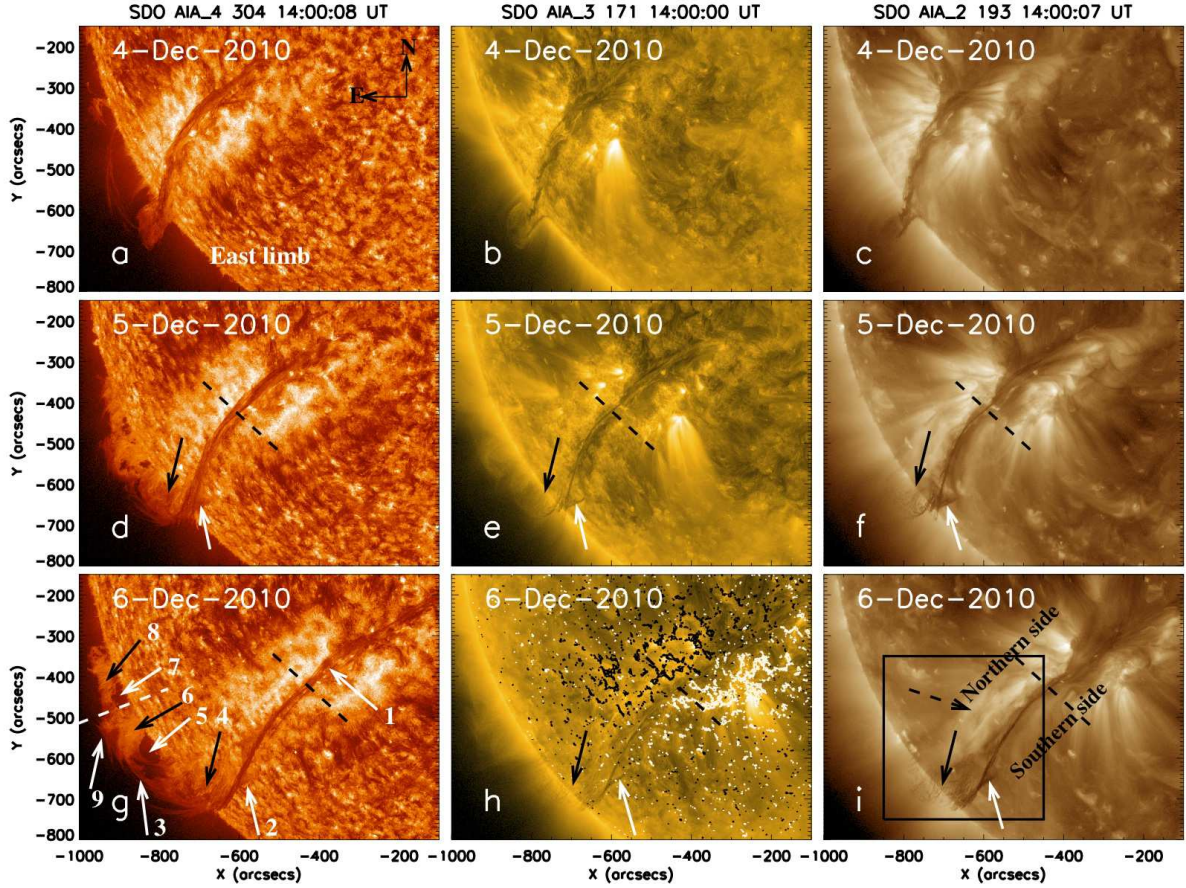}     
\end{center}
\caption{SDO/AIA observations of the quiescent prominence at 14:00 UT on 2010 December 3 (top row),  December 4 (middle row), and December 5 (bottom row). The white and black contours refer to the positive and negative polarities as observed by SDO/HMI at 14:00 UT on 2010 December 6. (A color version of this figure is available in the online journal.) AIA observations of the prominence at 193~\AA~and 304~\AA~ within 24 hours before the eruption are also available (video 1) in the electronic edition of the \emph{Astrophysical Journal}. The field of view of the online video contains only the prominence near the limb. The images are also rotated for 120 degree in the clockwise direction. \label{fig2}}
\end{figure}

Figures \ref{fig1}--\ref{fig2} shows SDO/AIA observations of the evolution of the prominence 
within six days (December 1--6) prior to the eruption. The images on the left, middle, and right columns are
taken at 304~\AA, 171~\AA, and 193~\AA, respectively. Note that artificial color is used for the on-line only figures.
Red color is used for the images taken at 304~\AA~by both AIA and EUVI, while 171~\AA~images by AIA are 
yellow. Images at 193~\AA~by AIA are dark yellow, while EUVI images at 195~\AA~are green. Moreover, both
193~\AA~and 195~\AA~provide similar spectral information. On December 1, the prominence is observed at the east limb and
contains mainly horizontal threads shown as emission at 304~\AA~(white arrows in Figure \ref{fig1}a).
A dark feature is also seen in absorption at all three wavelengths (black arrows in Figures \ref{fig1}a--\ref{fig1}c).
On December 2 and 3, the first part of the prominence rotated onto the disk, and the prominence at the limb
becomes more complex. One can see some dark vertical threads with overlying bright horizontal threads  
as shown in Figures 1d and 1g. Images at 304~\AA~shows the vertical threads either in emission or absorption. 
The dark vertical threads are better seen at  171~\AA~and 193~\AA.
Sometimes, bright emission in the prominence is also visible at 171~\AA~(e.g., Figure \ref{fig1}h). This bright emission in 171~\AA~is analyzed by \citet{2012arXiv1205.5460P}. Figure \ref{fig2} shows the prominence from December 4 to December 6. The filament on the disk shown at 171~\AA~and 193~\AA~appears to be much narrower than the one shown at 304~\AA. SDO/HMI observations (black and white contours in Figure \ref{fig2}h) show that the underlying magnetic fields are positive on the southern side and negative on the northern side of the filament. 

The AIA observations suggest that this prominence can be divided into three parts:
a region with mostly horizontal threads (Part I), a region with horizontal threads overlying the vertical threads (Part II), and a region with 
mostly vertical threads (Part III). This prominence contains several different features which are marked as numbers in Figure \ref{fig2}g.
Part I is located in the active region, which is on the northwestern side of the black dashed line in Figure \ref{fig2}g. The horizontal threads (marked as ``1" in Figure \ref{fig2}g) in Part I are best viewed at the limb in Figure \ref{fig1}a. Part II is located between the black and white dashed lines as shown in Figure \ref{fig2}g. The limb view of this region is shown in the images from December 2 to December 6 in Figures \ref{fig1}--\ref{fig2}.  Feature ``2" is a dense column composed of dark vertical threads which appear to be perpendicular to the overlying horizontal threads. The limb view of Part II displays an arch-like structure. The two feet of this arch (marked as ``4" and ``6") contain mostly quasi vertical threads, while the top of the arch is composed of mostly horizontal threads (feature ``3"). Note that the threads in feature 4 are not vertical, and the prominence spine appears to tilt towards the South. Feature 5 refers to the hole located below the top of the arch. Part III is located on the northern side of the white dashed line. This part is composed of mostly vertical threads (feature ``8") overlying a bubble-like structure (feature ``7"). Feature ``9" refers to a horn-like structure located above the vertical threads. All prominence structures are clearly visible at 304~\AA, while only parts of the prominence are seen as absorption in 171~\AA~and 193~\AA, which best reveal the fine structure of the vertical threads. Note that some features are also visible as emission in 171~\AA.

The aforementioned AIA observations reveal the filament while it rotating from the east limb to the disk.  STEREO$\_$B
views this filament as it rotating from the disk to the west limb. Figures \ref{fig3}--\ref{fig4} show the filament from December 1 to December 6
at 304~\AA~and 195~\AA~observed by EUVI aboard STEREO$\_$B. On December 1, the filament displays a dark long and narrow structure. The filament on the east appears to grow wider and thicker with time. The black dashed line in Figure \ref{fig4} divides the filament into two parts. The western part of the filament (active region part) contains mainly horizontal threads, while the eastern part (quiet Sun part) contains mainly quasi-vertical threads. Unlike AIA, EUVI observations do not reveal the horizontal threads located on top of the quasi-vertical threads. A comparison of Figures \ref{fig4}c and \ref{fig4}e shows that the active region part of the filament appears to rise up on December~6.  

Figure \ref{fig5} shows the filament observed at H$\alpha$ by KSO within four different days prior the eruption. H$\alpha$ observations show similar filament structures as those found at the EUV channels by AIA. A comparison of Figures \ref{fig5}c and \ref{fig5}d shows that most part of the active region filament (marked as black arrows) disappeared before the eruption on December 6. This is consistent with the rise of the active region filament as observed by EUVI.

The prominence before the eruption is very dynamic. Horizontal counter-streaming along the top of the arch as shown in Figure \ref{fig2}g is clearly seen in the AIA 304~\AA~observations (see video 1). Downflows are also observed in part of the vertical threads. Part of the vertical threads are just waving around without clear flows as shown in the AIA 193~\AA~observations. EUVI observations (video 2) show that the filament material is ejected from the active region part to the quiet Sun part. This ejection leads to the horizontal oscillations of the vertical threads and the formation of a dark tree-like structure (marked using white arrows in Figure~\ref{fig4}) near the border between the active region and the quiet Sun.  

\subsubsection{Dense Column Structure}

Figure \ref{fig6} shows comparisons of the AIA and EUVI observations of the filament on December 6. The field of view of the AIA and EUVI images are shown as black boxes in Figure \ref{fig2}i and Figure \ref{fig4}f, respectively. Figures \ref{fig1}--\ref{fig2} and Figures \ref{fig3}-\ref{fig4} suggest that the appearance of the filament is very different in AIA and EUVI observations. Therefore, to compare different filament features observed by the two instruments, we first identify a series of bright-point like features, which are marked as numbers in Figure \ref{fig6}. A careful comparison shows that the dense column observed by AIA  corresponds to the tree-like structure observed by EUVI. This dense column is formed by the strong accumulation of quasi-vertical threads (marked as white arrows in all images).

\subsubsection{The Structure around the Prominence}

\begin{figure}
\begin{center}
\epsscale{1.} \plotone{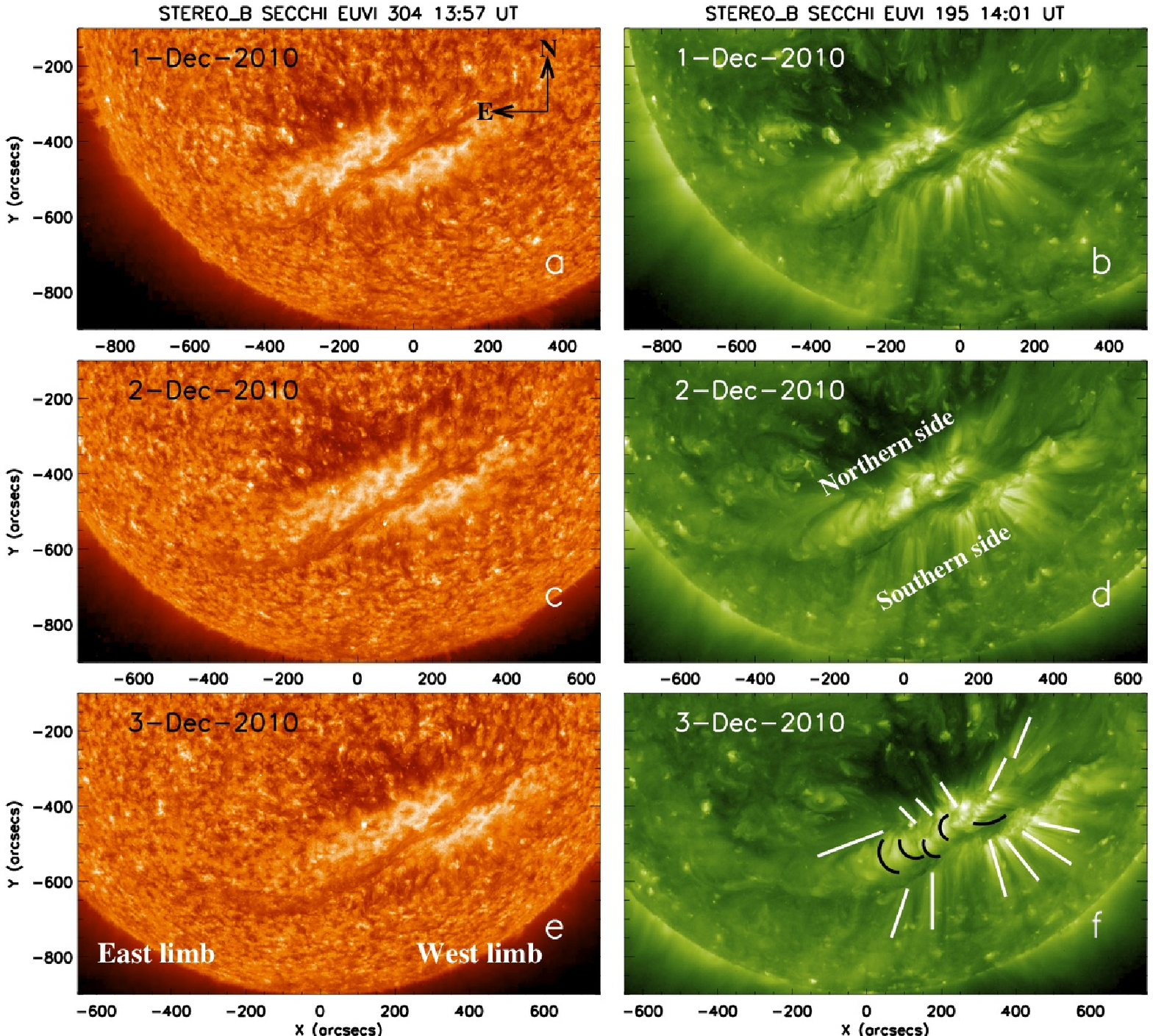}     
\end{center}
\caption{STEREO$\_$BB EUVI observations of the quiescent prominence around 14:00 UT on 2010 December 1 (top row),  December 2 (middle row), and December 3 (bottom row). (A color version of this figure is available in the online journal.) \label{fig3}}
\end{figure}

\begin{figure}
\begin{center}
\epsscale{1.} \plotone{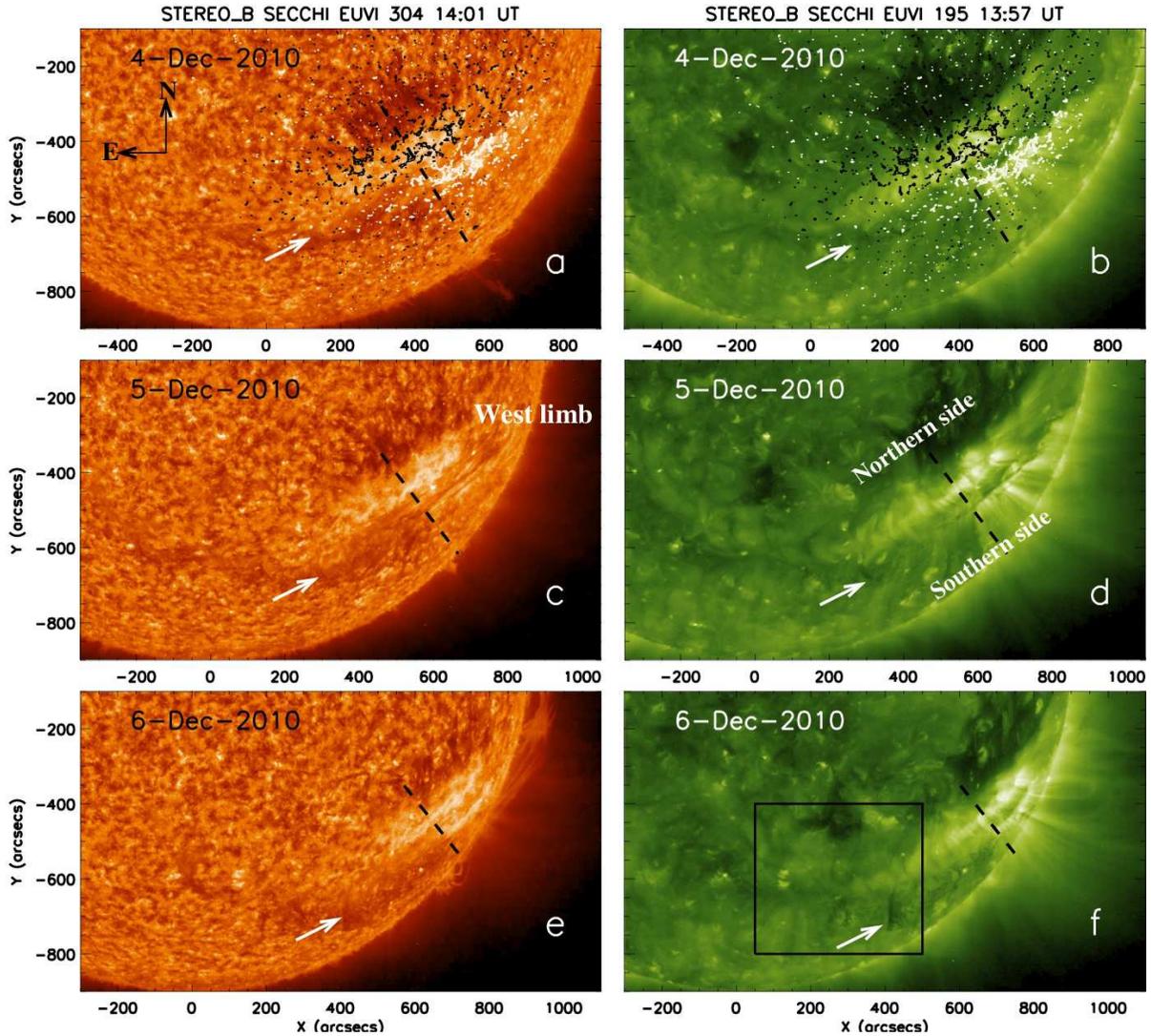}     
\end{center}
\caption{STEREO$\_$B/EUVI observations of the quiescent prominence around 14:00 UT on 2010 December 4 (top row),  December 5(middle row), and December 6 (bottom row). The white and black contours refer to the positive and negative polarities as observed by SDO/HMI at 14:00 UT on 2010 December 6.  (A color version of this figure is available in the online journal.) STEREO$\_$B observations of the prominence at 193~\AA~and 304~\AA~within 24 hours before the eruption are also available (video 2) in the electronic edition of the \emph{Astrophysical Journal}. \label{fig4}}
\end{figure}

\begin{figure}
\begin{center}
\epsscale{0.8} \plotone{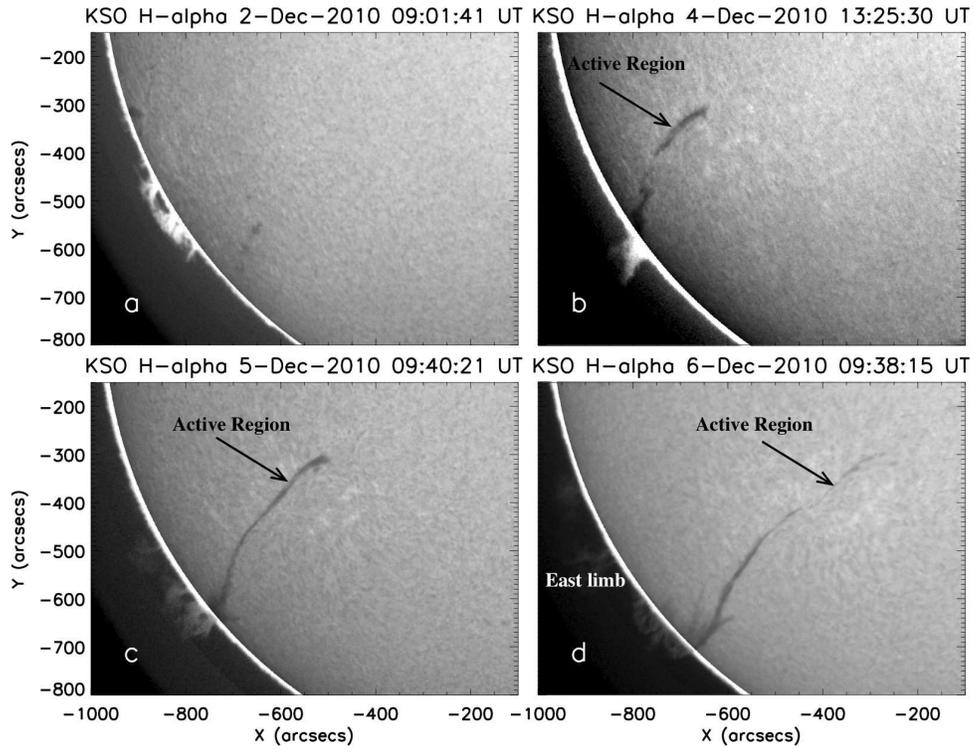}     
\end{center}
\caption{KSO H$\alpha$ observations of the quiescent prominence from December 2 to December 6. \label{fig5}}
\end{figure}

\begin{figure} 
\begin{center}
\epsscale{1.0} \plotone{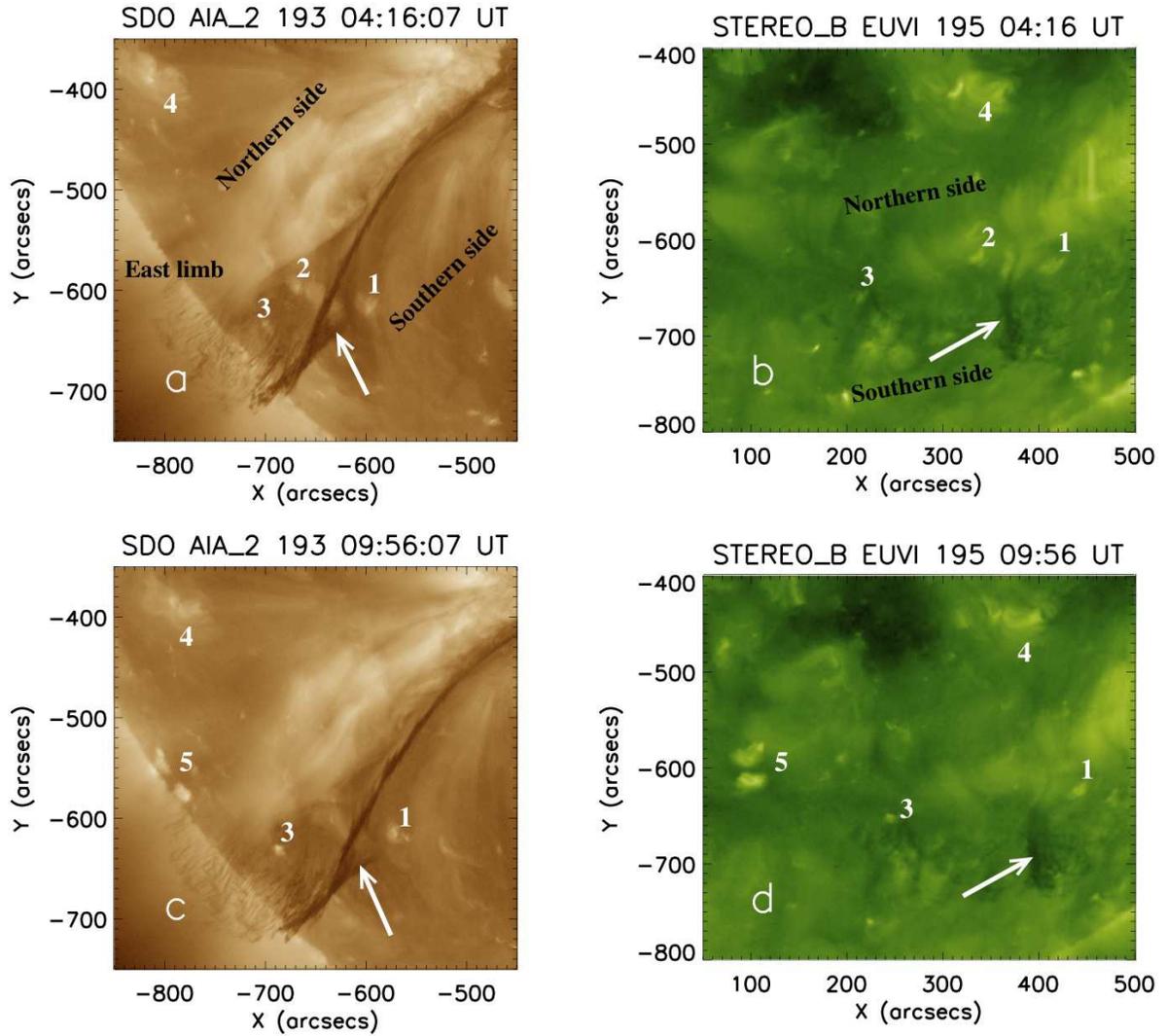}     
\end{center}
\caption{SDO/AIA (left column, 193~\AA) and STEREO$\_$B/EUVI (right column, 195~\AA) observations of the dense column structure on 2010 December 6 prior to the eruption. (A color version of this figure is available in the online journal.) \label{fig6}}
\end{figure}

\begin{figure} 
\begin{center}
\epsscale{0.8} \plotone{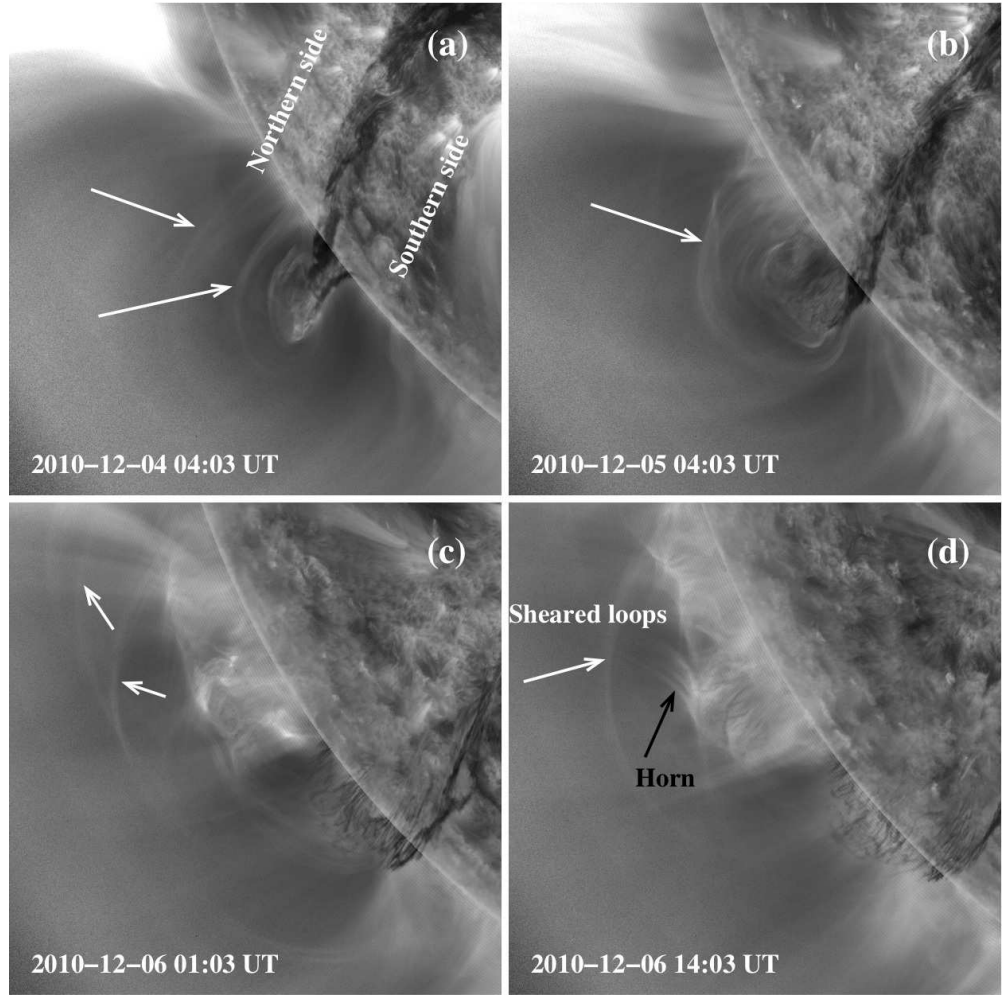}     
\end{center}
\caption{SDO/AIA 171~\AA~observations of loops surrounding the prominence prior to the eruption. Each image is the average of 30 images taken within 6 minutes. All images are contrast enhanced using a radial filter technique developed by S. Cranmer (2010, private communication). 
The field of view of each image is 540$^{\prime\prime}\times$540$^{\prime\prime}$. An animation of this figure (video 3) is available in the online journal. Note that radial filter technique is not applied for the animation.\label{fig7}}
\end{figure}

The prominence is located in a very narrow filament channel, which is best observed at 195~\AA~by EUVI as shown in Figure \ref{fig3}.
Straight faint features (traced using white lines in Figure \ref{fig3}f) are observed on the two sides of the channel. 
 These features may refer to the lower legs of the overlying coronal arcades.
Near the quiet Sun part, one can see some bright curved features (traced using black lines in Figure \ref{fig3}f),
which are located on the northern side of the filament channel. No clear counterparts of
these features can be identified on the southern side of the channel. This emission asymmetry that was first identified by \citet{2010ApJ...721..901S} is also
clearly visible in AIA observations, see Figures \ref{fig2}i and \ref{fig6}. However, these bright curved features appear to be 
straight in the AIA view. To have a better understanding on this asymmetry, we trace these features back when they are observed at
the east limb by AIA. Figure \ref{fig7} shows AIA 171~\AA~images on December 4--6. Each image is the average of 30 images taken within 6 minutes. All images are contrast enhanced using a radial filter technique developed by S. Cranmer (2010, private communication). 
The top images show bright loops which start from the northern side of the filament channel (white arrows), arch over the filament, and finally disappear behind the filament or join in the filament. Evolution of this type of loops from December 3 to December 6 are shown in the online video (video 3). 
The existence of these loops is consistent with the emission asymmetry in the sense that they do not have clear footpoints on the southern side of the filament channel. \citet{2010ApJ...721..901S} interpreted these bright curved features as magnetic field lines that turn into the flux rope.

The bottom images in Figure \ref{fig7} show highly sheared bright loops (white arrows) that go above the prominence. However, it is not clear where these loops end. Figure \ref{fig7}d also shows several bright horn-like features (black arrows) located right above the vertical prominence threads. This observation appears to suggest that some part of the flux rope lies above the prominence vertical threads.

\begin{figure} 
\begin{center}
\epsscale{0.7} \plotone{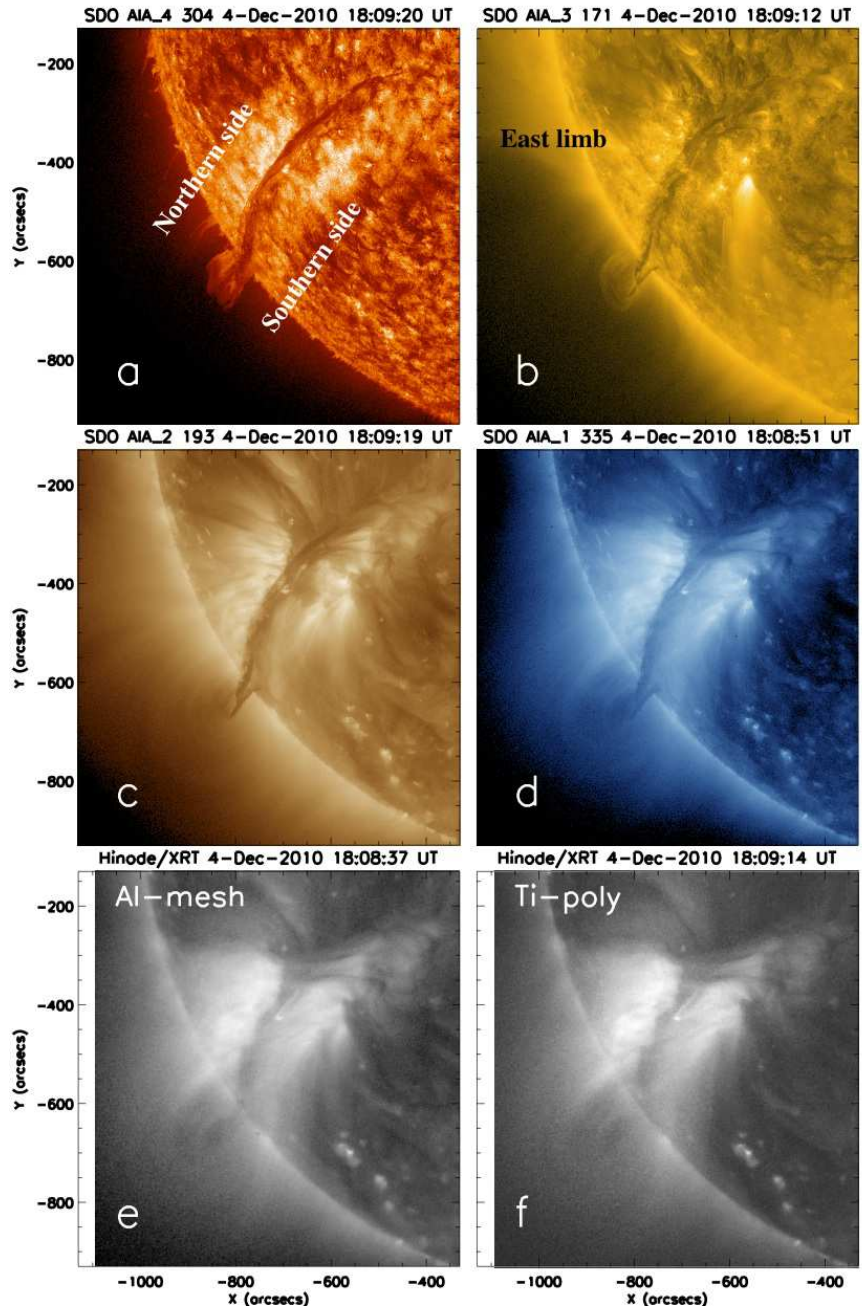}     
\end{center}
\caption{Multi-channel observations of the cavity structure around 18:09 UT on 2010 December 4. The images in the top two rows are provided by SDO/AIA, while Hinode/XRT images are shown in the bottom row. (A color version of this figure is available in the online journal.) \label{fig8}}
\end{figure}

Figure \ref{fig8} shows images of the prominence at different channels observed by SDO/AIA and Hinode/XRT on 2010 December 4.
A clear cavity surrounding the prominence can be seen in the two XRT images. The AIA image at 335~\AA~which is an average over 4 minutes shows similar structure as observed by XRT. No clear cavity can be identified in the other three EUV channels, i.e., 304~\AA, 171~\AA, and 193~\AA.
The AIA image at 335~\AA~(Figure \ref{fig8}d) shows that the dark filament lies on the southern edge of the cavity rather than at the center. We also can see that the bright coronal arcades surrounding the cavity appear to be asymmetric, i.e., the bright structure on the northern side is brighter than that on the southern side. This asymmetry is also clearly visible in the two XRT images (Figures \ref{fig8}e-- \ref{fig8}f). 

\begin{figure} 
\begin{center}
\epsscale{0.5} \plotone{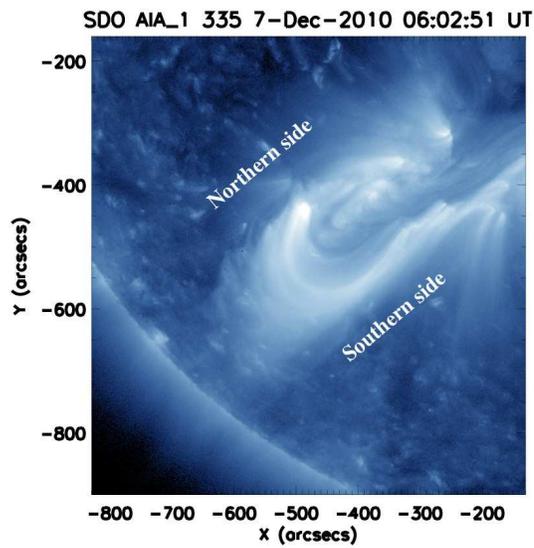}     
\end{center}
\caption{SDO/AIA (left column) of the post-eruption loops at 335~\AA~on 2010 December 7. (A color version of this figure is available in the online journal.) \label{fig9}}
\end{figure}

\citet{1996mpsa.conf..497M} found that there is a one-to-one correspondence between the dextral/sinistral orientation of the filament channel and the skew of the overlying coronal arcade. This implies that the axial magnetic field within filaments points in the same direction as that of the surrounding corona.  As viewed from the positive polarity side, for a sinistral filament the axial component of the field in the arcade points to the left, and vice versa \citep[i.e., Figure 1 in][]{2002SoPh..211..155M}.  Figure \ref{fig9} shows the post-eruption arcade above the erupted filament observed at 335~\AA~by AIA on 2010 December 7. As shown in Figure \ref{fig2}h, the positive polarity (white patches) is located on the southern side of the filament channel. This orientation of the arcade suggests that this filament is a sinistral filament, according to the aforementioned definition.

\subsubsection{Prominence Structure after the Eruption}

\begin{figure} 
\begin{center}
\epsscale{0.8} \plotone{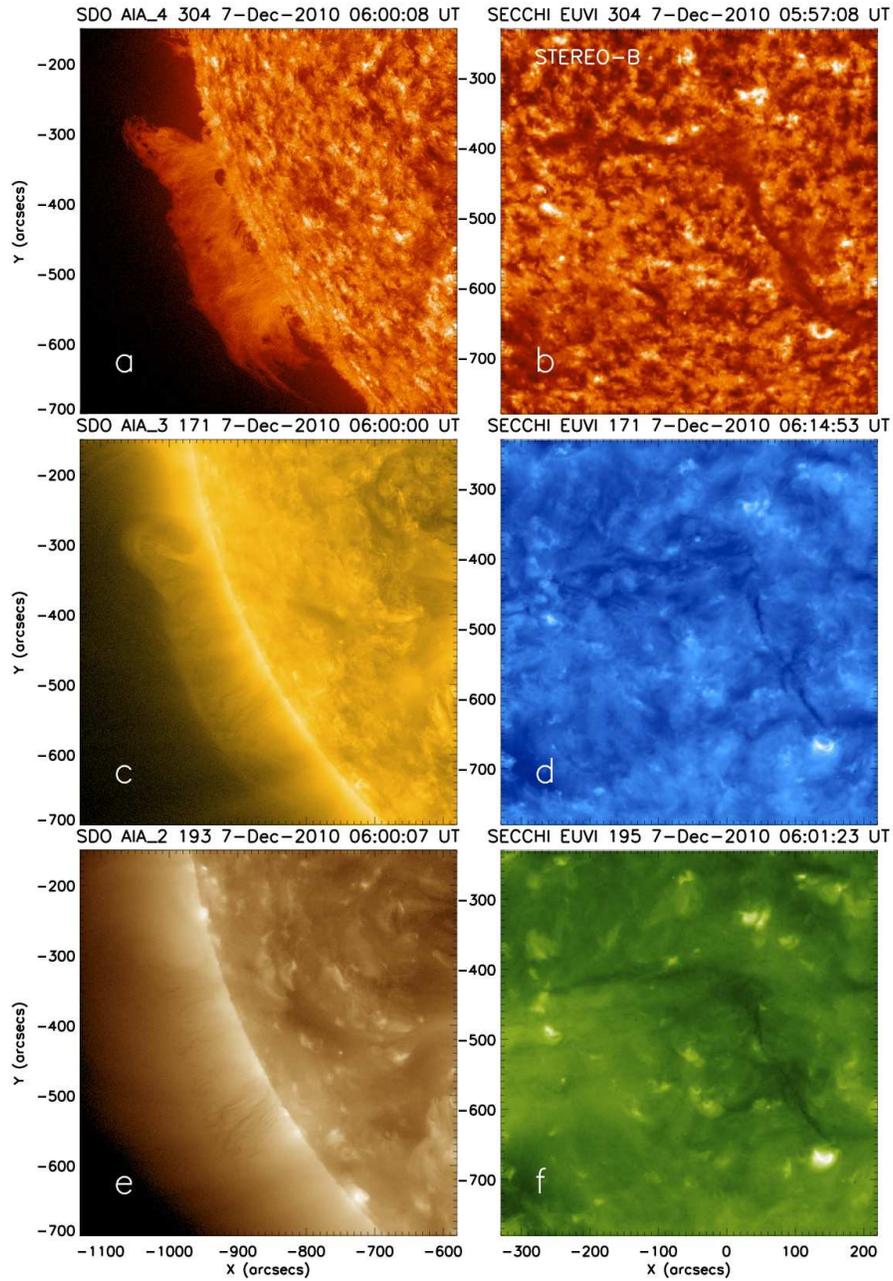}     
\end{center}
\caption{SDO/AIA (left column) and STEREO$\_$B/EUVI (right column) observations of the prominence after the eruption. The images at the top, middle, and bottom rows are taken at 304~\AA, 171~\AA, and 193/195~\AA~on 2010 December 7, respectively. (A color version of this figure is available in the online journal.) \label{fig10}}
\end{figure}

The prominence eruption on December 6 is a partial eruption, which means that part of the filament survived the eruption. Figure \ref{fig10} shows the east end of the quiescent prominence after the eruption. SDO observes this prominence at the east limb (left column), while STEREO$\_$B views it on the disk (right column). The side view of this prominence by AIA suggests that it contains mainly vertical threads. Images at 304~\AA~shows the vertical threads either in emission or absorption. Dark quasi-vertical threads are seen in absorption at both 171~\AA~and 193~\AA. In 171~\AA, one can also see a bright horizontal structure located on top of the dark vertical threads. The filament displays a thin dark structure, when viewing it from the top as shown in the STEREO$\_$B observations. The width of the filament at 304~\AA~(Figure \ref{fig10}b) is larger than that at the other two EUV channels (Figures \ref{fig10}d--\ref{fig10}e). Without the different point of view from SDO, it is very difficult for us to imagine that this thin dark filament contains vertical threads. It looks very similar to the active region filament which contains mainly horizontal threads. This thin dark filament structure may be the horizontal filament threads located on top of the vertical threads. Another possibility is that it is just the accumulation of the lined up dark vertical threads, and there are no horizontal threads. The spatial resolution of EUVI does not allow us to distinguish whether there are overlying horizontal threads or not.

\section{Models for the Observed Prominence}

It has been suggested that Quiescent Prominences (QPs) are located in
helical flux ropes that lie horizontally above filament channels on
the quiet Sun \citep[e.g., ][]{1994SoPh..155...69R,1995ApJ...443..818L}. 
The prominence plasma is assumed to be located
at dips in the helical field lines. In this section we construct
three-dimensional (3D) flux rope models for the observed prominence,
then we compare the models with AIA and EUVI observations.
The coronal arcade overlying a QP flux rope is thought to play an
important role in its equilibrium and stability. To have stable
equilibrium, the magnetic tension of the overlying arcade must balance
the magnetic pressure of the flux rope. If the overlying fields are
too weak, the flux rope will rise and expand into the heliosphere.
Therefore, an important open question is whether QP flux ropes can
exist in stable equilibrium with their surroundings. The overlying
fields are anchored in the quiet photosphere, which often has mixed
polarity with only a small excess of one polarity over the other.
It is unclear how effective such mixed-polarity fields can be
in anchoring QP flux ropes. The purpose of the present modeling is
to determine whether the observed QP can be explained in terms of a
stable flux rope model.

\subsection{Methodology}
\label{sect:method}

In this subsection we describe how the models are constructed.
In the last few years, we have developed various tools for modeling
non-potential magnetic fields in the solar corona, called the Coronal
Modeling System (CMS). Its main purpose is to construct NLFFF models of the coronal magnetic field.
The models are constructed by inserting a magnetic flux rope into a potential-field model of the
region and then applying magneto-frictional relaxation (MFR).  This
``flux rope insertion'' method is quite flexible and provides
information about the stability of the resulting magnetic fields
\citep[][]{2011ApJ...734...53S}. In the present study we apply this
flux rope insertion method to a quiescent prominence for the first time.

The observed QP is more than a solar radius in length, and the
model needs to include the overlying coronal arcade, so a large
simulation domain with spherical geometry is required. However,
we also need to resolve the mixed-polarity fields on the quiet Sun,
which requires high spatial resolution (a few Mm or better).
To achieve these goals, the CMS code uses variable grid spacing:
at low heights the grid spacing is about $0.002 \cos \lambda$
$\rm R_\odot$, where $\lambda$ is the latitude, but at larger heights
the cell size is increased by powers of 2. The high-resolution
(HIRES) computational domain extends about $117^\circ$ in longitude,
$48^\circ$ in latitude, and up to $2.41$ $\rm R_\odot$ from Sun
center. The magnetic field ${\bf B} ({\bf r})$ in this domain
is described in terms of vector potentials, ${\bf B} = \nabla \times
{\bf A}$.

The lower boundary condition for the HIRES region is derived from line-of-sight
(LOS) photospheric magnetograms obtained with the SDO/HMI. Since the prominence is observed near the east limb,
we have to use magnetograms that are taken several days after the
prominence eruption on December 6. We combine four magnetograms taken on 2010
December 8, 9, 10 and 11 (each at 14:00 UT) to construct a
high-resolution map of the radial component $B_r$ of magnetic field as
function of longitude and latitude at the lower boundary of the HIRES
region. We also use a SOLIS synoptic map of $B_r$ to compute a
low-resolution global potential field, which provides the side
boundary conditions for the HIRES domain, and also allows us to trace
field lines that pass through the side boundaries of the HIRES region.

\begin{figure}[ht!]
\epsscale{0.8}
\plotone{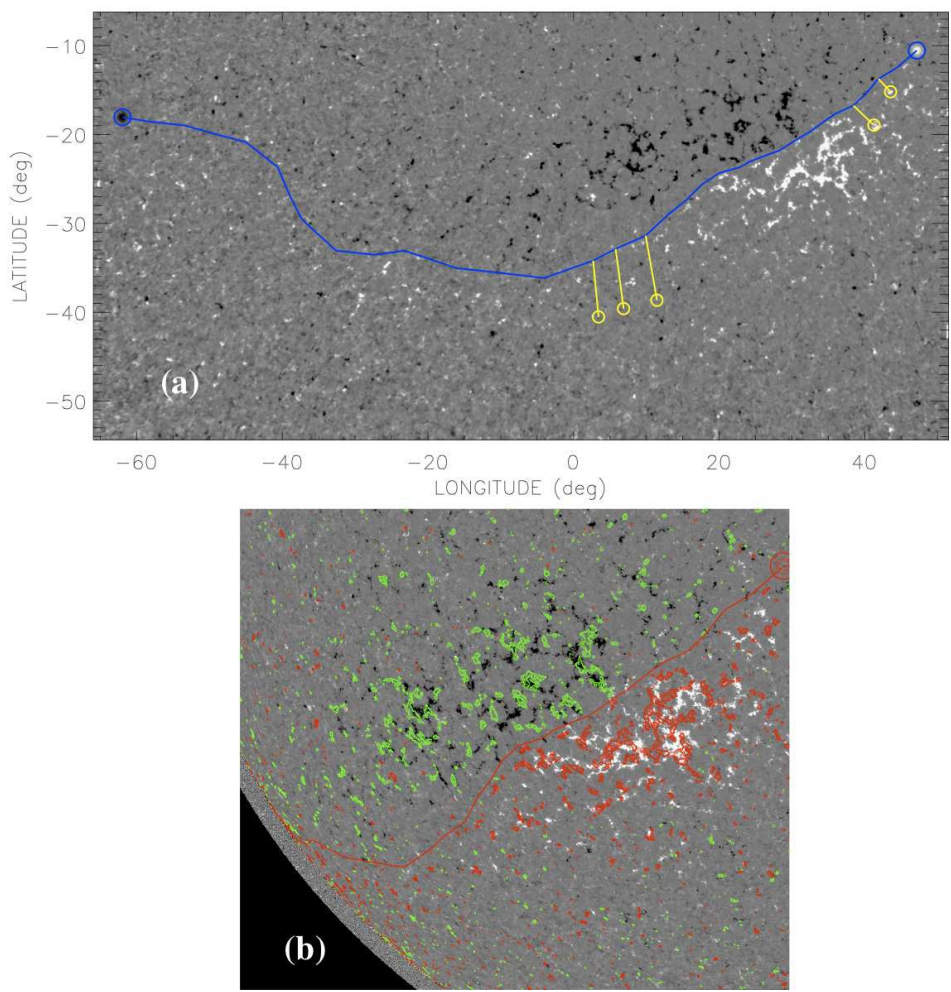}
\caption{(a) Longitude-latitude map of the radial component of magnetic
field in photosphere in the HIRES region of the model. The zero-point of
the longitude corresponds to the central meridian on 2010 December
10 at 14:00 UT. The blue curve shows the path along which the flux
rope is inserted into the model. Extra axial fluxes are added to (right) or removing from (left)
the flux rope along places indicated by the yellow bars. (b) The magnetic map shown in (a) (red and green contours)
is overlaid on the HMI magnetogram taken at 14:00 UT on 2012 December 6 (black and white image). The path along which the flux rope is inserted 
is shown as red curve in this image. Note that there is no systematic change in the flux distribution from December 6 to December 10. \label{fig11}}
\end{figure}

The HIRES magnetic map is shown in Figure \ref{fig11}a. Note that
outside the active region the field has mixed polarity with
dominantly positive polarity on the south side of the PIL and negative
polarity on the north side. Based on this map alone it is difficult to
recognize exactly where the PIL is located. Therefore, we used
co-aligned AIA images and HMI magnetograms to locate the base of the
prominence/filament on the magnetic map (blue curve in Figure
\ref{fig11}). This curve is the path along which the flux rope will
be inserted into the model. At the two ends of the path (blue
circles) the flux rope is anchored in the photosphere. 
Figure \ref{fig11}b shows the zoomed-in view of magnetic map shown in Figure \ref{fig11}a (red and green 
contours) overlaid on the HMI LOS magnetogram taken at 14:00 UT on December 6
(black and white image). This figure shows that there is no significant 
differences between the magnetic fields taken before the eruption (on December 6) and
after the eruption (December 10 and 11). One may notice some shift in individual  
magnetic patches, but no significant flux emergence and cancellation can be identified.
Moreover, the polarity inversion line appears to remain the same. It suggests that 
using HMI data several days after the eruption to model the prominence just before
the eruption is reasonable.

The methodology of flux rope insertion has been described elsewhere
\citep[e.g.,][]{2008ApJ...672.1209B, 2009ApJ...691..105S}, so only
a brief summary will be given. First, the potential field is
computed from the HIRES and global magnetic maps. Then by appropriate
modifications of the vector potentials a ``cavity'' is created above
the selected path, and a thin flux bundle (representing the axial flux
of the flux rope) is inserted into the cavity. Circular loops are
added around the flux bundle to represent the poloidal flux of the
flux rope. Then MFR is applied to drive the magnetic field toward a
force-free state. MFR is basically an evolution of the magnetic field
according to the magnetic induction equation with the velocity
taken to be proportional to the Lorentz force \citep[e.g.,][]
{1986ApJ...309..383Y, 2000ApJ...539..983V}. The magnetic diffusion
is kept as small as possible, so that the magnetic topology of the
field is approximately conserved (some diffusion is needed to prevent
numerical artifacts). If the axial and/or poloidal fluxes of the flux
rope are too large, the model will not reach an equilibrium state but
the field will keep expanding and moving to large heights, similar to
what happens on the real Sun during an eruption. In this paper we are
interested only in stable configurations, so we seek values of the
model parameters that produce stable NLFFFs. In previous work we
usually iterated the induction equation about 30,000 times, but we
found that for QP flux ropes this is not enough to reach an
equilibrium state. For the models presented here we use 90,000
iterations. 

For the models discussed in this paper, two magnetic sources with
fluxes of $\pm 2 \times 10^{20}$ Mx were added to the $B_r$-map at the
two ends of the filament path in order to enhance the stability of the
flux rope. The axial flux near the ends of the (blue) path equals the
flux of the sources ($\Phi_{\rm axi} = 2 \times 10^{20}$ Mx). The
axial field points to the left (east) as seen by an observer looking
at the QP from the south polar region, which has positive
polarity. Therefore, the flux rope has {\it sinistral} orientation,
consistent with the observed skew of the post-eruption arcades (see
Figure \ref{fig9}) and with the hemispheric pattern of chirality of filament
channels \citep{1994ssm..work..303M}.

The observations clearly show a filament within the active region,
and the STEREO$\_$B observations provide constraints on its height.
Preliminary modeling showed that in order to reproduce the observed
active region filament the axial flux of the flux rope must exceed a
certain minimum value: $\Phi_{\rm axi} > 6 \times 10^{20}$ Mx.
However, such values were found to be too high for the eastern part of
the filament overlying the quiet Sun (longitude $< 0$ in Figure
\ref{fig11}). Axial fluxes in excess of $4 \times 10^{20}$ Mx lead
to the eruption of the eastern part as the surrounding quiet-Sun
fields are too weak to anchor the flux rope to the photosphere.
Therefore, we must consider models in which the axial flux
$\Phi_{\rm axi}$ varies with position along the flux rope. This is
achieved by adding magnetic flux to (or removing flux from) the flux
rope at certain places along the filament path. These places are
indicated by the yellow ``barbs'' in Figure \ref{fig11}. In fact,
real filaments have barbs that may indeed be sites where magnetic
flux enters or leaves the main filament \citep[e.g.,][]{2011SSRv..158..237L}. In Figure \ref{fig11},
the circle at the end of a yellow line segment indicates the source
(or sink) of the added flux. As we follow the (blue) filament path
from right to left in the figure, the two barbs near longitude
$+40^\circ$ each add $3 \times 10^{20}$ Mx to the flux rope, and the
three barbs near longitude $+5^\circ$ each subtract $2 \times 10^{20}$
Mx. Therefore, the axial flux in the active region (in between the two sets
of barbs) is $8 \times 10^{20}$ Mx, and the axial flux on the quiet
Sun (outside the two sets of barbs) is $2 \times 10^{20}$ Mx. The
barbs near $+5^\circ$ are put on the south side of the PIL in order to
force the turned-out field lines to arch over the flux rope on their
way to the negative polarities on the north side. We find that this
initial setup leads to stable models that are in rough agreement with
the observations.

\subsection{Modeling Results and Comparison with Observations}
\label{sect:results}

\begin{figure}[ht!]
\epsscale{0.8}
\plotone{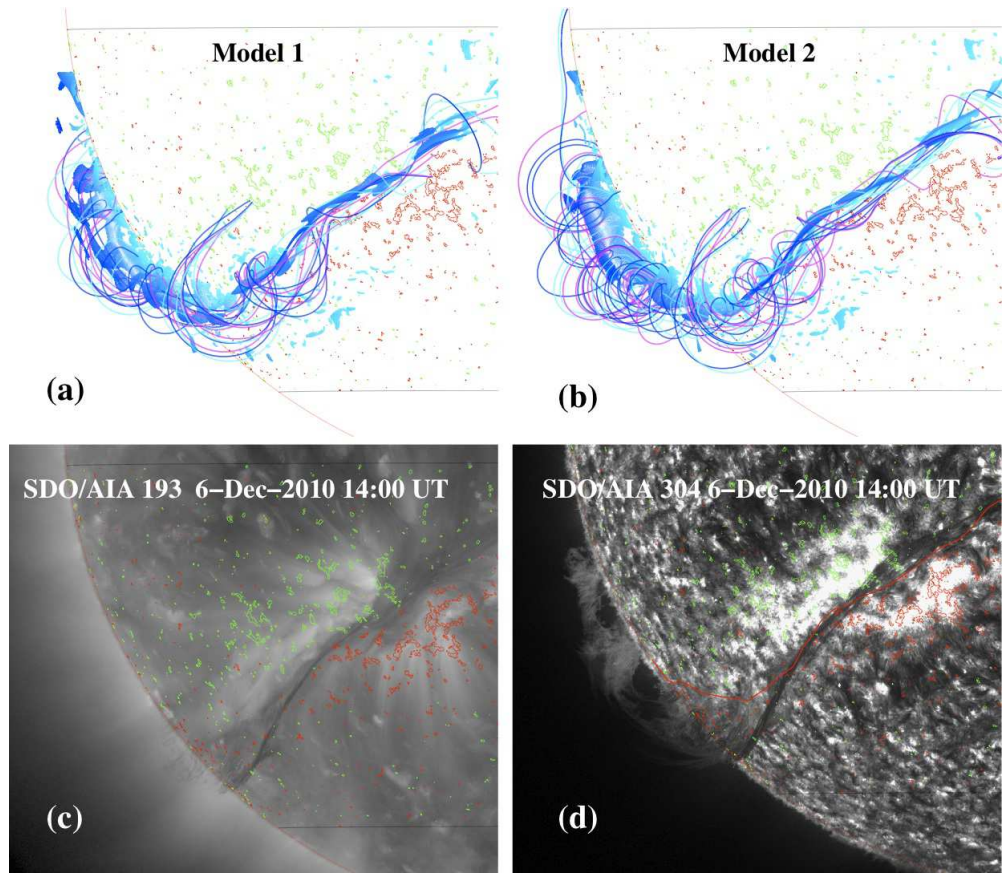}
\caption{Three-dimensional magnetic models of a quiescent prominence
observed with SDO/AIA on 2010 December 6 at 14:00 UT. The top panels
show results from two NLFFF models constructed using the flux rope
insertion method. For the model in panel (a) the axial flux of the
flux rope $\Phi_{\rm axi} = 2 \times 10^{20}$ Mx, and the initial poloidal
flux $F_{pol} = 1 \times 10^{10}$ $\rm Mx ~ cm^{-1}$, which produces a
relatively small degree of twist. Panel (b) shows a model with the
same axial flux but larger twist ($F_{\rm pol} = 2 \times 10^{10}$
$\rm Mx ~ cm^{-1}$). The colored curves are magnetic field lines in
and near the flux rope, the blue features indicate field-line dips,
and the red/green contours show the photospheric flux distribution
($B_r = \pm 71$ G, respectively). The red arc is the east solar limb,
and the black lines are the latitudinal boundaries of the HIRES
domain.  These features are projected onto the plane of the sky as seen
from the SDO spacecraft at 14:00 UT. Panels (c) and (d) show AIA images
taken in the 193 {\AA} and 304 {\AA} passbands. The path along which
the flux rope is inserted is shown as red curve in (d). Note that just inside
the limb the real prominence is displaced to the south with respect to
the PIL. This displacement is not reproduced in the NLFFF models.
\label{fig12}}
\end{figure}

In this paper we focus on models representing possible magnetic
configurations just prior to the eruption that started at 14:20 UT on
2010 December 6 (as discussed in section \ref{sect:method}, the models
are actually based on magnetograms taken several days later).
Two values for the poloidal flux $F_{\rm pol}$ of the flux rope are
considered: $1 \times 10^{10}$ $\rm Mx ~ cm^{-1}$ (Model 1), and $2
\times 10^{10}$ $\rm Mx ~ cm^{-1}$ (Model 2).  Figure \ref{fig12} shows results
from the NLFFF models and a comparison with AIA observation. The
top panels show the two models, projected onto the plane of the sky as
seen from SDO. The bottom panels show co-aligned images of the Sun
taken with AIA in the 193 {\AA} and 304 {\AA} passbands. The colored
curves in the top panels are magnetic field lines traced through the
model, and the blue features indicate the locations of dips in the
field lines, i.e., sites where the field lines are locally horizontal
and curved upward (light blue for low-lying dips, darker blue at
larger heights). The red and green contours show the magnetic flux
distribution on the photosphere. Note that the field lines in Figure
\ref{fig12}b are more twisted than those in Figure \ref{fig12}a,
which is due to the difference in the poloidal flux of the inserted
flux rope. Both models reproduce more or less the observed prominence
height (about 70 Mm) as seen in the AIA 304 {\AA} image (see Figure
\ref{fig12}d). This means that the modeled flux rope is sufficiently
thick, and the axis of the flux rope lies at sufficiently large height
to provide a reasonable model for the observed prominence everywhere
along its length. In the remainder of this paper we only consider the
model with poloidal flux $F_{\rm pol} = 10^{10}$ $\rm Mx ~ cm^{-1}$, i.e,
Model 1.

\begin{figure}[ht!]
\epsscale{0.5}
\plotone{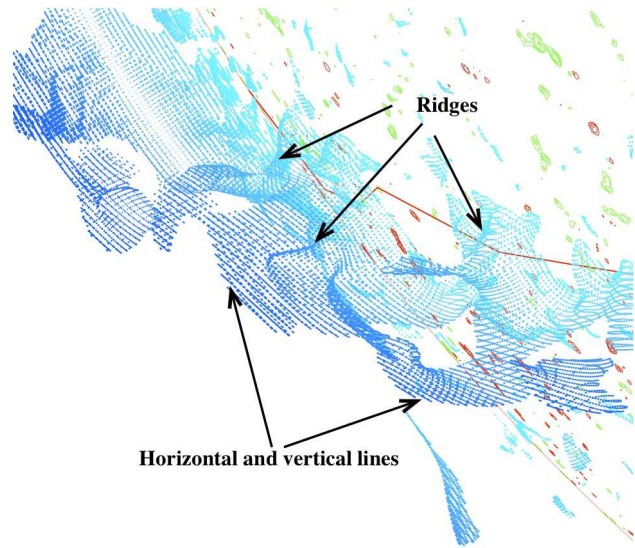}
\caption{Zoomed in view of the field line dips from Model 1 as shown in Figure \ref{fig12}a 
near the east limb (red arc). The red and green contours refer to the photospheric flux distribution. 
The red curve refers to the path along which the flux rope is inserted.
\label{fig13}}
\end{figure}

Figure \ref{fig13}  shows a zoomed-in view of the
middle part of Figure \ref{fig12}a. At this resolution the blue features resolve
into small blue dots that indicate the locations of field-line dips.
The dots lie on surfaces where the radial component of magnetic field
vanishes, $B_r (r,\theta,\phi) = 0$, where $r$, $\theta$ and $\phi$
are spherical coordinates. Note that the dots in Figure \ref{fig13} make
patterns of horizontal and quasi-vertical lines. These patterns should
not be taken too seriously because they are an artifact of the way
the dips are chosen. Specifically, we locate the intersections of the
surface $B_r = 0$ with the ``edges" of the 3D grid on which the
magnetic field ${\bf B}$ is defined (actually, the dots are  short
line segments indicating the direction of the horizontal field at the
dips, but this is not visible in Figure \ref{fig13}). Therefore, the patterns
of horizontal and vertical lines are created by interference with the
3D grid. However, we also see ``ridges" in Figure 13 where the density
of dots is higher than elsewhere. This indicates that the surface of
dips is strongly warped, and in some places our line-of-sight is
tangential to this warped surface. The warping of the $B_r = 0$
surface is due to the uneven distribution of magnetic sources on the
photosphere \citep[e.g.,][]{2003A&A...402..769A}. It has been suggested
that the quasi-vertical threads in quiescent prominences may be due
to such ridges \citep[e.g.,][]{2006ApJ...643L..65H, 2008SoPh..248...29D}.
However, we find that in our model the ridges are not always vertical,
and the number of ridges per unit length along the prominence is
far too small for the ridges to be identified with the observed
vertical threads, which usually are densely packed. Therefore, the
distribution of field-line dips in our NLFFF model does not reproduce
the observed quasi-vertical threads.

\begin{figure}[ht!]
\epsscale{0.8}
\plotone{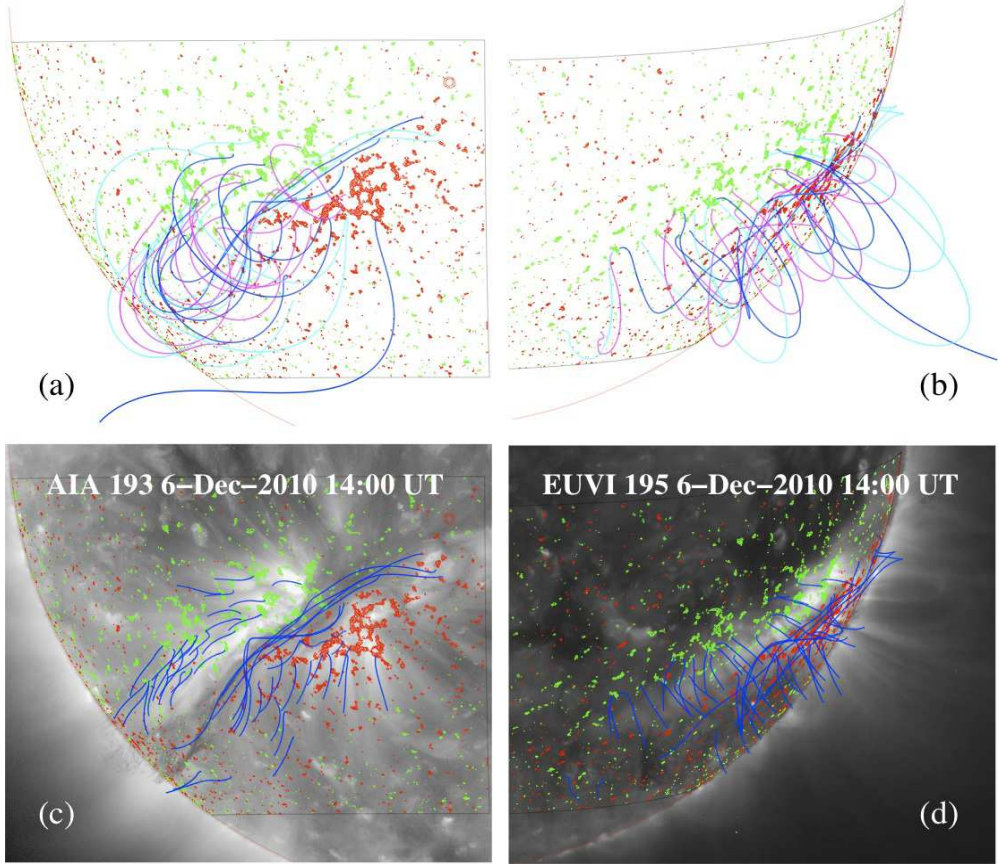}
\caption{Comparison of the model with SDO/AIA and STEREO$\_$BB/EUVI
observations of the coronal arcade surrounding the prominence cavity
(model with poloidal flux $F_{\rm pol} = 10^{10}$ $\rm Mx ~ cm^{-1}$).
The top panels show the same set of field lines (colored curves)
seen from two different viewpoints: (a) SDO and (b) STEREO$\_$B on 2010
December 6 at 14:00 UT. The red and green contours show the
photospheric flux distribution ($B_r = \pm 71$ G, respectively), and
the red arcs indicate the east and west limbs. The bottom panels show
images in (c) AIA 193 {\AA} and (d) EUVI 195 {\AA}, which are
dominated by Fe~XII 195 {\AA} emission. The images show thin threads
and fans that presumably are aligned with the local magnetic field.
On the quiet Sun the emissivity in the Fe~XII line drops off rapidly
with height (due to gravitational stratification of the plasma), so
only the plasma on the lower parts of the field lines is observable.
Therefore, we overplot the lower parts of the field lines that lie at
heights $h < 0.07$ $\rm R_\sun$ (blue line segments). Note that the
orientation of these line segments is consistent with the observed
fine structures, indicating the model reproduces the fields in the
coronal arcade.
\label{fig14}}
\end{figure}

Figure \ref{fig14} shows a comparison of Model 1 with AIA and EUVI
observations of the coronal arcade surrounding the prominence.
The top panels show the same set of field lines seen from two
different viewing angles: (a) from SDO, and (b) from STEREO$\_$B.
The bottom panels show images in AIA 193 {\AA} and EUVI 195 {\AA},
which are dominated by emission in Fe~XII 195 {\AA}. These images
show thin threads and fans that presumably are aligned with the local
magnetic field. On the quiet Sun the plasma density drops rapidly
with height due to the gravitational stratification of plasma along
the magnetic field lines. The Fe~XII ion is formed at temperatures of
about 1.5 MK, and the density scale height for such plasma in the low
corona is about 75 Mm. The emissivity in Fe~XII 195 {\AA} is
proportional to density squared, which decreases even more rapidly
with height. Therefore, we expect that in observations on the solar
disk the Fe~XII 195 {\AA} emission will be dominated by plasma in the
lower parts of coronal loops. Accordingly, in Figures \ref{fig14}c--\ref{fig14}d 
we overplot the lower parts of the field lines as blue line segments shown in \ref{fig14}a--\ref{fig14}b;
only heights less than $0.07$ $\rm R_\sun$ (49 Mm) are shown. Close inspection of these
images shows that the orientation of the line segments is roughly in
agreement with the directions of the thread-like structures seen by
AIA and EUVI. In particular, the model reproduces the bright features on the northern side of the PIL. 
Some of these threads are located on field
lines that turn into the flux rope as we follow its path from east
to west. This comparison between model and observations indicates
that Model 1 reproduces the surrounding magnetic fields quite well.
This model also confirms our previous interpretation of the emission asymmetry
on the two sides of the quiescent filament channels. Those bright
curved features on the north side of the PIL represent the lower part of the magnetic field lines 
that turn into the flux rope  \citep{2010ApJ...721..901S}.

\begin{figure}[ht!]
\epsscale{0.6}
\plotone{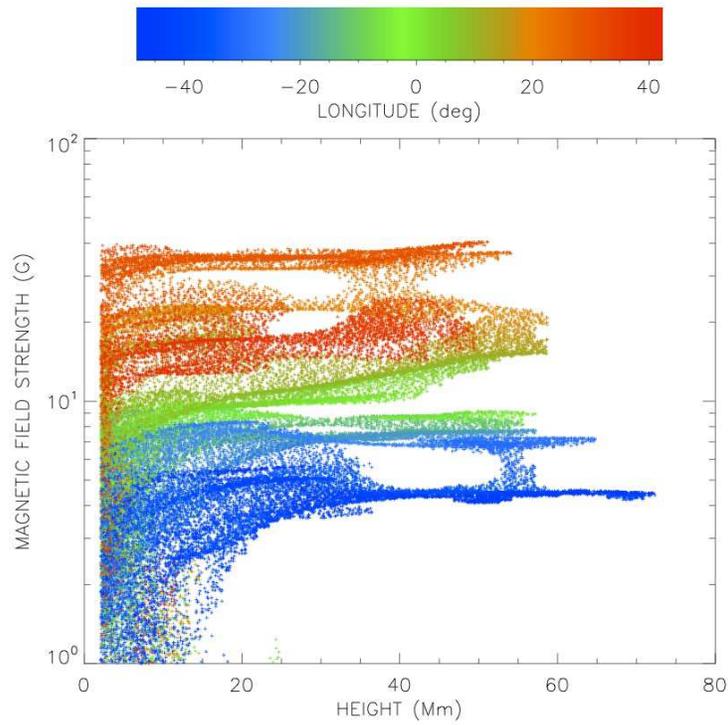}
\caption{Magnetic field strength at dips in the field lines from Model 1 are plotted 
as function of the height of the dips above the photosphere. The prominence plasma is assumed to be located
as such dips. All positions along the filament channel are included in this plot, except the two ends of the flux
rope where the rope is anchored in the photosphere. Different colors represent the field strength at different longitudes
as shown in Figure \ref{fig11}a.
\label{fig15}}
\end{figure}

Figure \ref{fig15} shows the magnetic field strengths
at dips in the field lines, plotted as function of the height $z$ of the
dips above the photosphere. The prominence plasma is assumed to be
located at such dips. All positions along the filament channel are
included in this plot, except the two ends of the flux rope where the
rope is anchored in the photosphere. The heights of the dips range up
to 70 Mm, and the field strengths range from 40 G in the active region
to 4.5 G in the quiescent part of the filament. The values of field strength are
consistent with measurements of prominence magnetic fields using the
Hanle effect \citep[e.g.,][]{1994SoPh..154..231B}, and with results from constant-$\alpha$
magnetohydrostatic modeling \citep{2003A&A...402..769A}. The plasma pressure in
quiescent prominences lies in the range 0.1 - 2.0 $\rm dyne ~ cm^{-2}$
\citep{1990LNP...363..298J, 2006ApJ...643L..65H}, so a magnetic field
strength of 4.5 G corresponds to a plasma $\beta$ in the range 0.12 -
2.5, where $\beta$ is the ratio of plasma- and magnetic pressures.
\citet{2006ApJ...643L..65H} have shown that when $\beta > 1$ the weight of
the prominence plasma will significantly distort the shape of the
magnetic field lines; such effects are not taken into account in the
present models. From the above estimates for $\beta$ it is clear that
gravity may significantly deepen the field-line dips in the quiescent
prominence compared to our NLFFF model. \citet{2006ApJ...643L..65H} have
developed 2D models of prominence threads supported by gravity-induced
dips in the field lines. In the active region the effects of gravity
are expected to be minor ($\beta < 0.03$).




There are interesting differences between our models and observations.
First, as mentioned above the AIA observations show thin vertical
threads in the prominence near the limb, but our models do not have
any such features in either the shapes of the magnetic field lines or
the distribution of field-line dips. This is not surprising because
our NLFFF models do not include the effects of plasma pressure and
gravity, which likely are responsible for producing the vertical
threads \citep{1957ZA.....43...36K, 2001A&A...375.1082H, 2005ApJ...626..551L, 2012ApJ...746..120H}.
Second, the model cannot explain the formation of the dense-column
structure.

\begin{figure}[ht!]
\epsscale{0.8}
\plotone{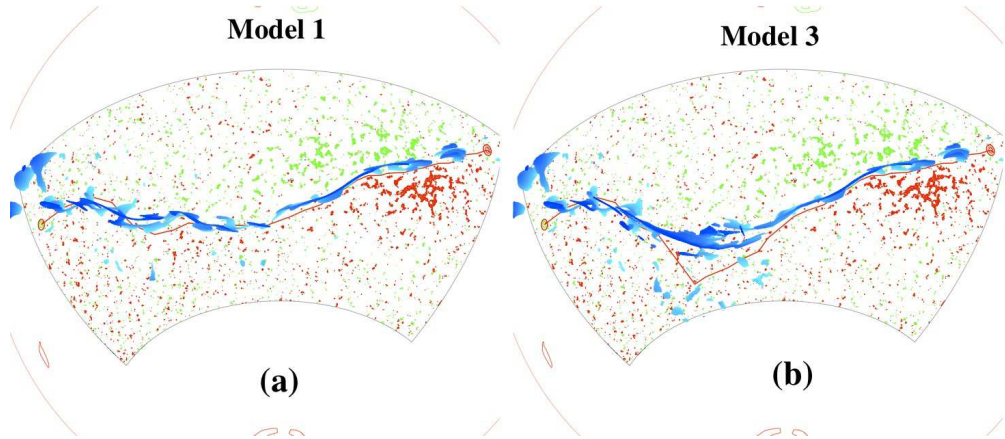}
\caption{Top view of the field line dips (blue features) from Model 1 and Model 3. Model 3 has the same parameters as Model 1 except the 
filament path (red curve). The red and green contours refer to the photospheric flux distribution.
\label{fig16}}
\end{figure}

The PIL is shown as a blue curve in Figure \ref{fig11}, and as a red curve in Figure \ref{fig12}d. Note that 
the filament spine is displaced to the South relative to the PIL in Figure \ref{fig12}d, whereas the base of the filament
appears to be located near the PIL. Therefore, the prominence does not lie vertically above the PIL, but
it tilted to the South. This tilt is not reproduced in either Model 1 or Model 2. In the models the flux
rope lies directly above the PIL, which is where the flux rope is inserted into the model. 
We create another model (Model 3) in order to see whether we can reproduce the tilt of the
prominence by changing the path where the flux rope is inserted. 
Figure \ref{fig16} shows top view of the field line dips (blue features) from Model 1 and Model 3. Model 3
has the same parameters as Model 1 except that the filament path (red curve) is displaced to the
south in the longitude range between $-35^{\circ}$ and $-5^{\circ}$.
The path in Model 3 is along the filament spine rather than the filament base 
as used in the other two models (see Figure \ref{fig12}d). Figure \ref{fig16}b suggests that 
the field line dips are pushed back to the PIL (filament base) after 90000-iteration MFR, though
the flux rope is originally inserted along the filament spine (red curve). Therefore, the tilt of the 
filament cannot be better reproduced by changing the path of the flux rope.
This absence of tilt is also not due to any lack
of convergence of the MFR; there simply is no force in the model to
push the flux rope sideways. 
Therefore, the observed tilt of the QP is
somewhat of a mystery. 

\begin{figure}[ht!]
\epsscale{0.8}
\plotone{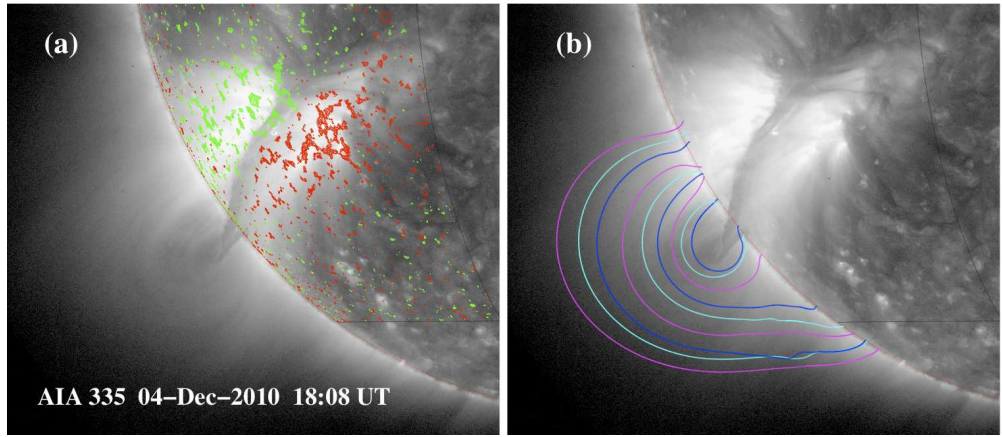}
\caption{Comparison of the observed cavity and magnetic field lines from Model 1. 
(a) SDO/AIA image of the cavity at 18:08 UT on 2010 December 4. The red and green
contours refer to photospheric magnetic fields observed by SDO/HMI. (b) The same image
as in (a) overlaid with magnetic field lines (color lines) surrounding the flux rope from Model 1.
\label{fig17}}
\end{figure}

Figure \ref{fig17} shows comparison of the observed cavity and magnetic field
lines from Model 1. The AIA image in Figure \ref{fig17}a shows 
that the cavity is tall and narrow  rather than a half circle-like shape.
The size and shape of the model flux rope is represented by the inner most dark
blue line shown in Figure \ref{fig17}b. A comparison of these two figures suggests
that the observed cavity is taller than the modeled flux rope.
The observations show that the bright features surrounding the cavity are
nearly vertical with respect to the solar limb. However, the lower
part of the modeled field lines on the south side of the cavity are 
much more inclined in comparison to the observations. 

The two models presented in Figure \ref{fig12} lie at the edge of the
stable regime. Indeed, an eruption does occur starting at 14:20 UT,
shortly after the time for which the models are constructed. However,
the eruption involves only the western part of the QP; the eastern
part remains behind and is apparently quite stable (see Figure \ref{fig10}).
This partial eruption is difficult to understand on the basis of our
models: the flux rope extends along the full length of the QP, and
therefore we expect that the eruption should remove the entire flux
rope. This leads us to question the validity of the flux rope model
for describing the magnetic structure of this QP.

\section{Discussion and Conclusions}

A large polar crown prominence partially erupted on 2010 December 6. To understand the magnetic support of the prominence, we focus on the structure and dynamics of the prominence before and after the eruption. The 6-day SDO/AIA observations of the filament near the east limb before the eruption suggest that this filament can be divided into 3 parts. The active region part near the west end contains mainly horizontal threads. The middle part near the quiet Sun contains vertical threads with overlying horizontal threads. While the east end of the prominence that survived the eruption appears to be composed of mainly vertical threads. The post-eruption limb observations by AIA suggest that the prominence left behind contains mainly vertical threads, while a regular thin dark filament structure appears in the on-disk observations by EUVI. This thin dark filament structure may be the horizontal filament threads located on top of the vertical threads. Another possibility is that it is just the accumulation of the lined up dark vertical threads, and there are no horizontal threads. Such structure, consisting of horizontally aligned threads, was reported by \citet{2004SoPh..221..297S} for another filament. Corresponding H$\alpha$ appearance of such threads was discussed by \citet{2006ApJ...643L..65H}.

STEREO/EUVI observes straight and faint features representing the lower legs of the overlying coronal arcades on the two sides of the filament channel. On the quiet Sun part, bright features with no clear counterparts are also identified on the northern side of the channel. This emission asymmetry is consistent with the existence of loops that originates from the northern side of the channel, then arch over the filament and disappear or join in the filament as shown in earlier limb observations at 171~\AA~by AIA. A clear cavity corresponding to the filament channel is observed on December 4 at 335~\AA~by AIA and XRT. The observations suggests that the filament is located on the southern edge of the cavity rather than at the center. The bright features (lower legs of the coronal arcades) on the north side of the cavity are brighter than those on the south side. The skew of the post-eruption arcade suggests that this filament is sinistral.

The prominence before the eruption is very dynamic. AIA observes horizontal counter-streaming overlying the vertical threads. STEREO$\_$B/EUVI observes filament material ejection from the active region part to the quiet Sun part. This ejection leads to strong oscillations of the vertical threads and the formation of a dense column near the border between the active region and quiet Sun. The dense column refers to strong accumulation of dark vertical filament threads.

Using the flux rope insertion method developed by \citet{2004ApJ...612..519V} we construct two non-linear force-free field models. The poloidal flux ($1\times 10^{20}$ Mx cm$^{-1}$) in Model 1 is half of that in Model 2, so the flux rope in Model 1 is less twisted. In both models, the axial fluxes are $8 \times 10^{20}$ Mx  and $2 \times 10^{20}$ Mx in the active region and quiet Sun, respectively. Since the prominence is observed at the east limb, the models are constructed based on the combined LOS photospheric magnetogram observed by SDO/HMI several day after the prominence eruption. Both models lie at the edge of the stable regime.

Our models match some of the observed features quiet well. For example, the height and location of the field line dips in our models appear to match those of the prominence threads, though the narrow vertical thread-like structure is not reproduced in our model. The lower part of the magnetic field lines from our models can replicate the bright features on the two side of the filament channel quite well. Our model also confirms our previous interpretation of the emission asymmetry on the two sides of the channel on the quiet Sun \citep{2010ApJ...721..901S}. This emission asymmetry is due to the fact that some of the bright features on the northern side (the brighter side) of the channel represent the field lines that turn into the flux rope. Therefore, no counterparts of these features can be found on the southern side of the channel. We suggest that plasma may be injected into the prominence along the aforementioned field lines that turn into the flux rope, since these field lines start from the photosphere. This is consistent with the thermal nonequilibrium mechanism for prominence formation as suggested by Antiochos, Karpen, and colleagues \citep[e.g., ][]{1991ApJ...378..372A, 2005ApJ...635.1319K, 2012ApJ...746...30L}. In this model, localized heating above the flux tube footpoints produces evaporation of the chromospheric plasma, which condenses in the coronal part of the flux tube.

Acknowledgments: We thank the anonymous referee for valuable comments to improve this paper. Hinode is a Japanese mission developed and launched by ISAS/JAXA, with NAOJ as domestic partner and NASA and STFC (UK) as international partners. It is operated by these agencies in co-operation with ESA and the NSC (Norway). We thank the team of SDO/AIA, SDO/HMI, STEREO/EUVI, Hinode/XRT, KSO, and SOLIS for providing the valuable data. The EUVI and HMI data are downloaded via the Virtual Solar Observatory and the Joint Science Operations Center. Y. Su acknowledges Dr. Mark Weber for helpful discussions and thanks Dr. Suli Ma for helping on the radial filter technique. This project is partially supported under contract NNM07AB07C from NASA to the Smithsonian Astrophysical Observatory (SAO) and SP02H1701R from LMSAL to SAO as well as NASA grant NNX12AI30G.

\clearpage


\begin{thebibliography}{}

\bibitem[Antiochos 
\& Klimchuk(1991)]{1991ApJ...378..372A} Antiochos, S.~K., \& Klimchuk, J.~A.\ 1991, \apj, 378, 372 

\bibitem[Antiochos et al.(1994)]{1994ApJ...420L..41A} Antiochos, S.~K., 
Dahlburg, R.~B., \& Klimchuk, J.~A.\ 1994, \apjl, 420, L41 

\bibitem[Asgari-Targhi 
\& van Ballegooijen(2012)]{2012ApJ...746...81A} Asgari-Targhi, M., \& van Ballegooijen, A.~A.\ 2012, \apj, 746, 81 

\bibitem[Aulanier 
\& D{\'e}moulin(2003)]{2003A&A...402..769A} Aulanier, G., \& D{\'e}moulin, P.\ 2003, \aap, 402, 769 

\bibitem[Aulanier et al.(2002)]{2002ApJ...567L..97A} Aulanier, G., DeVore, 
C.~R., \& Antiochos, S.~K.\ 2002, \apjl, 567, L97 

\bibitem[Aulanier et al.(1998)]{1998A&A...335..309A} Aulanier, G., Demoulin, P., van Driel-Gesztelyi, L., Mein, P., \& Deforest, C.\ 1998, \aap, 335, 309 

\bibitem[Berger et al.(2008)]{2008ApJ...676L..89B} Berger, T.~E., Shine, 
R.~A., Slater, G.~L., et al.\ 2008, \apjl, 676, L89 

\bibitem[Berger et al.(2010)]{2010ApJ...716.1288B} Berger, T.~E., Slater, 
G., Hurlburt, N., et al.\ 2010, \apj, 716, 1288 

\bibitem[Bobra et al.(2008)]{2008ApJ...672.1209B} Bobra, M.~G., van Ballegooijen, A.~A., \& DeLuca, E.~E.\ 2008, \apj, 672, 1209

\bibitem[Bommier et al.(1994)]{1994SoPh..154..231B} Bommier, V., Landi 
Degl'Innocenti, E., Leroy, J.-L., 
\& Sahal-Brechot, S.\ 1994, \solphys, 154, 231 

\bibitem[Brueckner et al.(1995)]{1995SoPh..162..357B} Brueckner, G.~E., et 
al.\ 1995, \solphys, 162, 357 

\bibitem[Canou \& Amari(2010)]{2010ApJ...715.1566C} Canou, A., \& Amari, T.\ 2010, \apj, 715, 1566 

\bibitem[Chae(2010)]{2010ApJ...714..618C} Chae, J.\ 2010, \apj, 714, 618 

\bibitem[Chae et al.(2008)]{2008ApJ...689L..73C} Chae, J., Ahn, K., Lim, 
E.-K., Choe, G.~S., \& Sakurai, T.\ 2008, \apjl, 689, L73 

\bibitem[Chae et al.(2001)]{2001ApJ...560..476C} Chae, J., Wang, H., Qiu, J., et al.\ 2001, \apj, 560, 476 

\bibitem[DeRosa et al.(2009)]{2009ApJ...696.1780D} DeRosa, M.~L., et al.\ 2009, \apj, 696, 1780 

\bibitem[DeVore 
\& Antiochos(2000)]{2000ApJ...539..954D} DeVore, C.~R., \& Antiochos, S.~K.\ 2000, \apj, 539, 954 


\bibitem[Dud{\'{\i}}k et al.(2008)]{2008SoPh..248...29D} Dud{\'{\i}}k, J., 
Aulanier, G., Schmieder, B., Bommier, V., 
\& Roudier, T.\ 2008, \solphys, 248, 29 

\bibitem[Engvold(1976)]{1976SoPh...49..283E} Engvold, O.\ 1976, \solphys, 
49, 283 

\bibitem[Foukal(1971)]{1971SoPh...19...59F} Foukal, P.\ 1971, \solphys, 19, 
59 

\bibitem[Gaizauskas(1998)]{1998ASPC..150..257G} Gaizauskas, V.\ 1998, IAU 
Colloq.~167: New Perspectives on Solar Prominences, 150, 257 

\bibitem[Gibson et al.(2006)]{2006ApJ...641..590G} Gibson, S.~E., Foster, 
D., Burkepile, J., de Toma, G., \& Stanger, A.\ 2006, \apj, 641, 590 

\bibitem[Golub et al.(2007)]{2007SoPh..243...63G} Golub, L., et al.\ 2007, 
\solphys, 243, 63 

\bibitem[Guo et al.(2010)]{2010ApJ...714..343G} Guo, Y., Schmieder, B., 
D{\'e}moulin, P., Wiegelmann, T., Aulanier, G., T{\"o}r{\"o}k, T., 
\& Bommier, V.\ 2010, \apj, 714, 343 

\bibitem[Heinzel 
\& Anzer(2001)]{2001A&A...375.1082H} Heinzel, P., \& Anzer, U.\ 2001, \aap, 375, 1082 

\bibitem[Heinzel 
\& Anzer(2006)]{2006ApJ...643L..65H} Heinzel, P., \& Anzer, U.\ 2006, \apjl, 643, L65 

\bibitem[Heinzel et al.(2008)]{2008ApJ...686.1383H} Heinzel, P., Schmieder, 
B., F{\'a}rn{\'{\i}}k, F., et al.\ 2008, \apj, 686, 1383 


\bibitem[Hillier et al.(2012)]{2012ApJ...746..120H} Hillier, A., Berger, 
T., Isobe, H., \& Shibata, K.\ 2012, \apj, 746, 120 

\bibitem[Hirayama(1985)]{1985SoPh..100..415H} Hirayama, T.\ 1985, \solphys, 
100, 415 

\bibitem[Howard et al.(2008)]{2008SSRv..136...67H} Howard, R.~A., et al.\ 
2008, \ssr, 136, 67 


\bibitem[Jensen 
\& Wiik(1990)]{1990LNP...363..298J} Jensen, E., \& Wiik, J.~E.\ 1990, IAU Colloq.~117: Dynamics of Quiescent Prominences, 363, 298 

\bibitem[Kano et al.(2008)]{2008SoPh..249..263K} Kano, R., et al.\ 2008, 
\solphys, 249, 263 

\bibitem[Karpen et al.(2005)]{2005ApJ...635.1319K} Karpen, J.~T., Tanner, 
S.~E.~M., Antiochos, S.~K., \& DeVore, C.~R.\ 2005, \apj, 635, 1319 

\bibitem[Kippenhahn 
\& Schl{\"u}ter(1957)]{1957ZA.....43...36K} Kippenhahn, R., \& Schl{\"u}ter, A.\ 1957, \zap, 43, 36  
  
\bibitem[Kosugi et al.(2007)]{2007SoPh..243....3K} Kosugi, T., et al.\ 
2007, \solphys, 243, 3   
  
\bibitem[Kucera et al.(2003)]{2003SoPh..212...81K} Kucera, T.~A., Tovar, 
M., \& de Pontieu, B.\ 2003, \solphys, 212, 81          
         
\bibitem[Kuperus \& Raadu(1974)]{1974A&A....31..189K} Kuperus, M., \& Raadu, M.~A.\ 1974, \aap, 31, 189

\bibitem[Labrosse et al.(2010)]{2010SSRv..151..243L} Labrosse, N., Heinzel, 
P., Vial, J.-C., et al.\ 2010, \ssr, 151, 243 

\bibitem[Lemen et al.(2012)]{2012SoPh..275...17L} Lemen, J.~R., Title, 
A.~M., Akin, D.~J., et al.\ 2012, \solphys, 275, 17 

\bibitem[Lin(2011)]{2011SSRv..158..237L} Lin, Y.\ 2011, \ssr, 158, 237 

\bibitem[Lin et al.(2003)]{2003SoPh..216..109L} Lin, Y., Engvold, O.~R., 
\& Wiik, J.~E.\ 2003, \solphys, 216, 109 

\bibitem[Lin et al.(2008)]{2008ASPC..383..235L} Lin, Y., Martin, S.~F., 
\& Engvold, O.\ 2008, Subsurface and Atmospheric Influences on Solar Activity, 383, 235 
 
\bibitem[Low \& Hundhausen(1995)]{1995ApJ...443..818L} Low, B.~C., \& Hundhausen, J.~R.\ 1995, \apj, 443, 818 

\bibitem[Low 
\& Petrie(2005)]{2005ApJ...626..551L} Low, B.~C., \& Petrie, G.~J.~D.\ 2005, \apj, 626, 551 

\bibitem[Luna et al.(2012)]{2012ApJ...746...30L} Luna, M., Karpen, J.~T., 
\& DeVore, C.~R.\ 2012, \apj, 746, 30 


\bibitem[Mackay et al.(2010)]{2010SSRv..151..333M} Mackay, D.~H., Karpen, 
J.~T., Ballester, J.~L., Schmieder, B., 
\& Aulanier, G.\ 2010, \ssr, 151, 333 

\bibitem[Martin et al.(1994)]{1994ssm..work..303M} Martin, S.~F., 
Bilimoria, R., \& Tracadas, P.~W.\ 1994, Solar Surface Magnetism, 303 

\bibitem[Martin(1998)]{1998SoPh..182..107M} Martin, S.~F.\ 1998, \solphys, 
182, 107 

\bibitem[Martin \& McAllister(1996)]{1996mpsa.conf..497M} Martin, S.~F., \& McAllister, A.~H.\ 1996, IAU Colloq.~153: Magnetodynamic Phenomena in the Solar Atmosphere - Prototypes of Stellar Magnetic Activity, 497 

\bibitem[McAllister et al.(2002)]{2002SoPh..211..155M} McAllister, A.~H., 
Mackay, D.~H., \& Martin, S.~F.\ 2002, \solphys, 211, 15

\bibitem[Menzel \& Wolbach(1960)]{1960S&T....20..252M} Menzel, D.~H., \& Wolbach, J.~G.\ 1960, \skytel, 20, 252 

\bibitem[Okamoto et al.(2007)]{2007Sci...318.1577O} Okamoto, T.~J., 
Tsuneta, S., Berger, T.~E., et al.\ 2007, Science, 318, 1577 

\bibitem[Parenti et al.(2012)]{2012arXiv1205.5460P} Parenti, S., Schmieder, 
B., Heinzel, P., \& Golub, L.\ 2012, arXiv:1205.5460 


\bibitem[Pneuman(1983)]{1983SoPh...88..219P} Pneuman, G.~W.\ 1983, 
\solphys, 88, 219 

\bibitem[Priest (1989)]{Priest1989}Priest, E.\ 1989, Dynamics and Structure of Quiescent Solar Prominences (Dordrecht: Kluwer) 

\bibitem[Priest et al.(1989)]{1989ApJ...344.1010P} Priest, E.~R., Hood, 
A.~W., \& Anzer, U.\ 1989, \apj, 344, 1010 


\bibitem[Rust \& Kumar(1994)]{1994SoPh..155...69R} Rust, D.~M., \& Kumar, A.\ 1994, \solphys, 155, 69 

\bibitem[R{\'e}gnier \& Priest(2007)]{2007A&A...468..701R} R{\'e}gnier, S., \& Priest, E.~R.\ 2007, \aap, 468, 701 

\bibitem[R{\'e}gnier et al.(2002)]{2002A&A...392.1119R} R{\'e}gnier, S., Amari, T., \& Kersal{\'e}, E.\ 2002, \aap, 392, 1119 

\bibitem[Savcheva \& van Ballegooijen(2009)]{2009ApJ...703.1766S} Savcheva, A., \& van Ballegooijen, A.\ 2009, \apj, 703, 1766 

\bibitem[Savcheva et al.(2012a)]{2012ApJ...744...78S}
Savcheva, A.S., \& van Ballegooijen, A.A., \& DeLuca, E.E. 2012, \apj,
744, 78

\bibitem[Savcheva et al.(2012b)]{2012ApJ...750...15S} Savcheva, A., Pariat, 
E., van Ballegooijen, A., Aulanier, G., \& DeLuca, E.\ 2012, \apj, 750, 15 


\bibitem[Schmieder et al.(2004)]{2004SoPh..221..297S} Schmieder, B., Lin, 
Y., Heinzel, P., \& Schwartz, P.\ 2004, \solphys, 221, 297 

\bibitem[Schou et al.(2012)]{2012SoPh..275..229S} Schou, J., Scherrer, 
P.~H., Bush, R.~I., et al.\ 2012, \solphys, 275, 229 

\bibitem[Su et al.(2011)]{2011ApJ...734...53S} Su, Y., Surges, V., van Ballegooijen, A., DeLuca, E., \& Golub, L.\ 2011, \apj, 734, 53 

\bibitem[Su et al.(2010)]{2010ApJ...721..901S} Su, Y., van Ballegooijen, A., \& Golub, L.\ 2010, \apj, 721, 901 

\bibitem[Su et al.(2009a)]{2009ApJ...704..341S} Su, Y., van Ballegooijen, A., Schmieder, B., Berlicki, A., Guo, Y., Golub, L., 
\& Huang, G.\ 2009, \apj, 704, 341

\bibitem[Su et al.(2009b)]{2009ApJ...691..105S} Su, Y., van Ballegooijen, A., Lites, B.~W., Deluca, E.~E., Golub, L., Grigis, P.~C., 
Huang, G., \& Ji, H.\ 2009, \apj, 691, 105 

\bibitem[Su \& van Ballegooijen(2012b)]{Su2012} Su, Y., van Ballegooijen, A.\ 2012, \apj, submitted

\bibitem[Thompson(2012)]{Thom2012} Thompson, W.T., \solphys, submitted

\bibitem[Tandberg-Hanssen (1995)]{Tandberg-Hanssen1995}Tandberg-Hanssen, E.\ 1995, The nature of solar prominences {Dordrecht: Kluwer}

\bibitem[van Ballegooijen(2004)]{2004ApJ...612..519V} van Ballegooijen, 
A.~A.\ 2004, \apj, 612, 519 

\bibitem[van Ballegooijen \& Cranmer(2010)]{2010ApJ...711..164V} van Ballegooijen, A.~A., \& Cranmer, S.~R.\ 2010, \apj, 711, 164 

\bibitem[van Ballegooijen \& Martens(1989)]{1989ApJ...343..971V} van Ballegooijen, A.~A., \& Martens, P.~C.~H.\ 1989, \apj, 343, 971 

\bibitem[van Ballegooijen et al.(2000)]{2000ApJ...539..983V} van 
Ballegooijen, A.~A., Priest, E.~R., \& Mackay, D.~H.\ 2000, \apj, 539, 983 

\bibitem[Wuelser et al.(2004)]{2004SPIE.5171..111W} Wuelser, J.-P., et al.\ 
2004, \procspie, 5171, 111 

\bibitem[Yang et al.(1986)]{1986ApJ...309..383Y} Yang, W.~H., Sturrock, 
P.~A., \& Antiochos, S.~K.\ 1986, \apj, 309, 383 

\bibitem[Zirker et al.(1998)]{1998Natur.396..440Z} Zirker, J.~B., Engvold, 
O., \& Martin, S.~F.\ 1998, \nat, 396, 440 

\end{thebibliography}
\end{document}